\documentclass[useAMS,usenatbib,a4]{mn2e}
\usepackage[english]{babel}
\usepackage[utf8]{inputenc}
\usepackage{makeidx}
\usepackage{array,multirow}
\usepackage{rotating}
\usepackage{color}
\usepackage{epsfig}
\usepackage{epstopdf}
\usepackage{float}
\restylefloat{table}
\usepackage{here}
\usepackage[font=small,format=plain,labelfont=bf,up,textfont=it,up]{caption}
\usepackage{graphicx}
\usepackage{verbatim}
\usepackage{amsmath}
\usepackage{multicol}
\usepackage{ulem}
\DeclareGraphicsExtensions{.pdf,.png,.jpg,.mps}

\renewcommand{\tabcolsep}{-1cm}

\title[Real-space reconstruction of void stacks]{Real-space density profile reconstruction of stacked voids}
\author[A. Pisani et al.]{A. Pisani$^{1,2}$\thanks{E-mail:
pisani@iap.fr (AP)
}, G. Lavaux $^{1,2,3,4,5}$, P. M. Sutter$^{1,2,6,7}$ and B. D. Wandelt$^{1,2,6}$
\\ \\
$^{1}$Sorbonne Universites, UPMC Univ Paris 06, UMR7095, Institut
d’Astrophysique de Paris, 98bis Bd. Arago, F-75014, Paris, France\\
$^{2}$CNRS, UMR7095, Institut d’Astrophysique de Paris, 98bis Bd. Arago, F-75014, Paris, France\\
$^{3}$Perimeter Institute for Theoretical Physics, 31 Caroline St. N., Waterloo, ON N2L 2Y5, Canada\\
$^{4}$Department of Physics and Astronomy, University of Waterloo, 200 University Avenue West, Waterloo, Ontario N2L 3G1, Canada\\
$^{5}$Canadian Institute for Theoretical Astrophysics, 60 St. George St., Toronto, ON M5S 3H8 Canada\\
$^{6}$Departments of Physics and Astronomy, University of Illinois at Urbana-Champaign, Urbana, IL 61801, USA\\
$^{7}$Center for Cosmology and Astro-Particle Physics, Ohio State University, Columbus, OH 43210, USA}

\begin{document}

\date{Accepted 2014 July 8. Received 2014 June 7; in original form 2013 June 12}

\pagerange{3238--3250} \pubyear{2014}

\maketitle

\label{firstpage}

\begin{abstract}

We present a non-parametric, model-independent method to reconstruct the spherical density profiles of void stacks in real space, \textit{without redshift-space distortions}. Our method uses the expected spherical symmetry of stacked voids to build the shape of the spherical density profile of cosmic voids in real space without any assumption about the cosmological model. We test the reconstruction algorithm with a toy model, a dark matter simulation and a mock galaxy catalogue. We present the result for the simulations: the reconstruction of the spherical density profile for simulated stacked voids in real space. We also present a first application of the algorithm to reconstruct real cosmic void stacks density profiles in real space from the Sloan Digital Sky Survey \citep{Sutter2012a}. We discuss capabilities of the algorithm and possible future improvements. Reconstructed density profiles from real voids open the way to the study of the spherically averaged dynamical structure of voids.
\end{abstract}


\section{Introduction}

In recent years, cosmologists developed an increasing interest in cosmic voids (for an historical review see \cite{Thompson2011} and \cite{Chincarini2013}). These structures shape the Universe at large scales as a cosmic web \citep{Bond1996}, along with filaments and clusters of galaxies. Voids, discovered in 1978 \citep{Gregory1978,Joeveer1978,Tully1978,Kirshner1981,deLapparent1986}, are under-dense regions in the Universe with sizes from ten to hundreds of Mpc. 

The appeal of cosmic voids is considerable: being nearly empty, they might be mainly composed of dark energy \citep{Bos2012}. Voids potentially are an important tool to study the effects of dark energy, but promise also to discriminate between different cosmological models (including modified gravity models such as fifth force models, as shown in \cite{Spolyar2013} and \cite{Clampitt2013}; or coupled dark matter-dark energy models, as discussed by \cite{Sutter2014}). The simplicity of the evolution of voids, compared to higher density zones of the Universe, is another asset in favour of their study.

Cosmic voids have, generally, very different shapes. 
But in a homogeneous and isotropic universe the average real-space shape of voids is spherical \citep{Ryden1996}; this feature is fundamental for our work. In such a universe there is no possible reason that could ever give to the void an average shape following preferred directions. The average shape of cosmic voids is obtained through stacking. The work of \cite{Lavaux2012}, based on numerical simulations and void stacking, suggests the existence of a general stacked profile of cosmic voids, roughly independent of void size and redshift. Real data of stacked voids \citep{Sutter2012a} from the Sloan Digital Sky Survey (SDSS) also seem to support the hypothesis of a common shape for the profile. Furthermore, the work of \cite{Hamaus2014} has investigated the existence of a simple empirical function to universally describe void profiles.

The density profile of a stacked cosmic void has a general shape with an underdensity on the centre; the density then increases towards its maximum value, reached at the over-dense \textit{wall} enclosing the void. The stacked wall consists in clumps, filaments and sheets. Outside the wall, the profile asymptotes to the mean density. The spherically symmetric density profile of the stacked void only depends on radius.

Redshift distortions affect the density profile of cosmic voids obtained until now (both in simulations and observations). To fully understand voids it is of crucial importance to recover the shape of the density profile \textit{without redshift distortions}. 

When observing galaxies in the Universe, we do not have real-space images. Surveys such as the SDSS measure the position in redshift space. Since our Universe is expanding, all galaxies are redshifted due to the expansion of space. To this is added the redshift caused by the peculiar motion of the galaxy. Only the line-of-sight component of velocity affects the galaxy redshift \citep{Hamilton1998}. In the framework of cosmic voids, this would mean that the real-space spherical shape of voids is \textit{distorted in redshift space} (as it emerges from both \cite{Lavaux2012} and \cite{Sutter2012b}).
 
If we consider only the study of the void itself, the \textit{peculiar velocities} of void galaxies are a measure of the evolution of the void. As a general behaviour, cosmic voids should flow out (as quantified by \cite{Patiri2012} and \cite{Aragon-Calvo2013}), with a motion of galaxies from the centre of the void towards the wall. The non linear part of peculiar velocities thickens the wall, \cite{Ceccarelli2006} studied the behaviour of velocities near the wall in mock catalogues (and in data, using the model of velocities obtained from simulations to analyse real voids). Generally, the effect of velocities is to increase the distortion of the void along the line-of-sight direction. 

The reconstruction of the spherical profile removes the effect of peculiar velocities and gives us the first real-space profiles of stacked voids. The reconstruction has two powerful assets: it does not make any assumption about the cosmological model or the physics of the void to get the real-space shape of voids (except for sphericity and an overall physical scale) and it does not need to model the peculiar velocity distortions to reconstruct the profile. 

This new possibility to determine the density profile of stacked voids in real space using the spherical symmetry opens the way to many applications. These include the study of dark energy and the constraint of cosmological parameters. Since dark energy should strongly rule the evolution of cosmic voids (where matter is rare), the physics of the voids is directly linked to dark energy (see \cite{Lee2009} and \cite{Bos2012}). The determination of the density profile of cosmic voids offers a promising avenue to probe their contents.

The reconstruction of the spherical density profile of cosmic voids promises also to improve the application of Alcock-Paczy\'{n}ski test (illustrated in \cite{Alcock1979}) to voids (first suggested by \cite{Ryden1995}, studied and applied in \cite{Lavaux2012} and \cite{Sutter2012b}). It is not the purpose of this paper to illustrate this method (see \cite{Sutter2012b}), we will give only a brief explanation to show the importance of a correct measure of the spherical density for its application. 

The Alcock-Paczy\'{n}ski test applied to cosmic voids compares the shape of the distorted void in redshift space and of the spherical void in real space (of course for stacked voids, otherwise sphericity could not be assumed) to obtain information about the expansion of the Universe; it uses the void as a standard sphere.

Since the distortion is a combined effect of the expansion of the Universe and of the peculiar velocities of galaxies, the knowledge of the spherical density profile of voids in real space would lead to a more precise application of the Alcock-Paczy\'{n}ski test to measure the expansion of the Universe. The determination of the density profile of stacked cosmic voids in real space is the first step to a model of the effect of peculiar motions and promises to improve the application of the test.

As pointed out by \cite{Verde2013}, in light of the recent results from the \textit{Planck} satellite (see \cite{Planck2013}) and of the tension risen with data from Type Ia supernovae \citep{Riess1998,Perlmutter1999}, a local cosmological-independent measure of the Hubble parameter (potentially accessible with the Alcock-Paczy\'{n}ski test) assumes great importance.  

The paper is organized as follows: in Section 2 we explain the method to recover the profile in real space, we present the algorithm for the reconstruction and we test it with a toy model of voids. In Section 3 we apply the method to a full dark matter simulation and obtain the shape of the spherical density profile of a simulated stacked void in real space, independently from the cosmological model. In Section 4 we further test the reconstruction algorithm on stacked voids obtained from a mock galaxy catalogue. In Section 5, we present a first application of the algorithm to stacked cosmic voids from SDSS data \citep{Sutter2012a} and we discuss capabilities of the algorithm. We finally conclude in Section 5 by a summary and discussion on future purposes for the use of the algorithm and possible improvements for further applications to data from real surveys. 

\section{Spherical density profile reconstruction: the method}

\subsection{General approach for a standard sphere}

For a large number of voids the stacked voids of \cite{Sutter2012a} can be considered standard spheres. Peculiar velocities and the expansion of the Universe distort the standard sphere in redshift space along the line-of-sight. The basic idea is that we would like to remove the distortion to reconstruct the spherical shape in real space. Our method uses the fact that the \textit{projection} of the void stack along the line-of-sight does not depend on redshift-space distortions.

If we are then able to reconstruct the sphere from the projection, we will have the spherical density profile in real space, that is without redshift distortions. We recall that the reconstructed density profile for a stacked void will simply be a function of the radius, since the void is spherically symmetric in real space. The idea is shown in Fig. \ref{method}. We note that this can be done for voids of reasonable size (smaller than 100 $\mathrm{h^{-1}Mpc}$) and at low redshift ($z\ll1$), where the angular distance is independent of redshift (at higher redshift some angular effects can appear, depending if the galaxy is in front of or behind the centre of the void).

In the next subsection we will briefly introduce redshift distortions and explain how they affect the shape of the void.

\subsection{Spherical density reconstruction}
In order to understand correctly how to recover the spherical profile, we need to give a description of redshift distortions.

\subsubsection{Redshift distortions}
For the purpose of this paper, we simply want to present the method to recover the density profile, study its feasibility and show a first application as a proof of concept. The analysis of redshift distortions is simplistic and we leave for future work a more detailed analysis. We consider approximations valid at low redshift ($z\ll1$) and low curvature for an isotropic and homogeneous universe.
The redshift distance is obtained considering the real distance plus the effect of peculiar velocities along the line-of-sight. 
Following the notation in \cite{Hamilton1998}, along the line-of-sight direction we have: $s=r+v \mathrm{cos} \theta$, where
 $s$ is the redshift distance in velocity units, equal to $cz$;
 $r$            is the true distance; and
 $v$      is the peculiar velocity, projected along the line-of-sight direction by defining the angle  $\theta$ between the line-of-sight direction and the velocity.
  We then have
 \begin{equation}
 cz=H_{0}d+v\mathrm{cos}\theta \label{vel}
 \end{equation}
  where
 $c$ is the speed of light, 
 $z$            is the redshift of the galaxy, 
 $H_{0}$    is Hubble constant, and
 $d$      is the distance of the galaxy. 
We will now define the distorted, projected and spherical densities necessary to apply the method.
 
\subsubsection{Distorted, projected and spherical densities} 
In this section, we define some notation useful to the discussion of the method. We consider the density of the void, where by density we mean the number of galaxies per volume element (a number density).

\begin{figure}
\centering \includegraphics[width=0.3\textwidth]{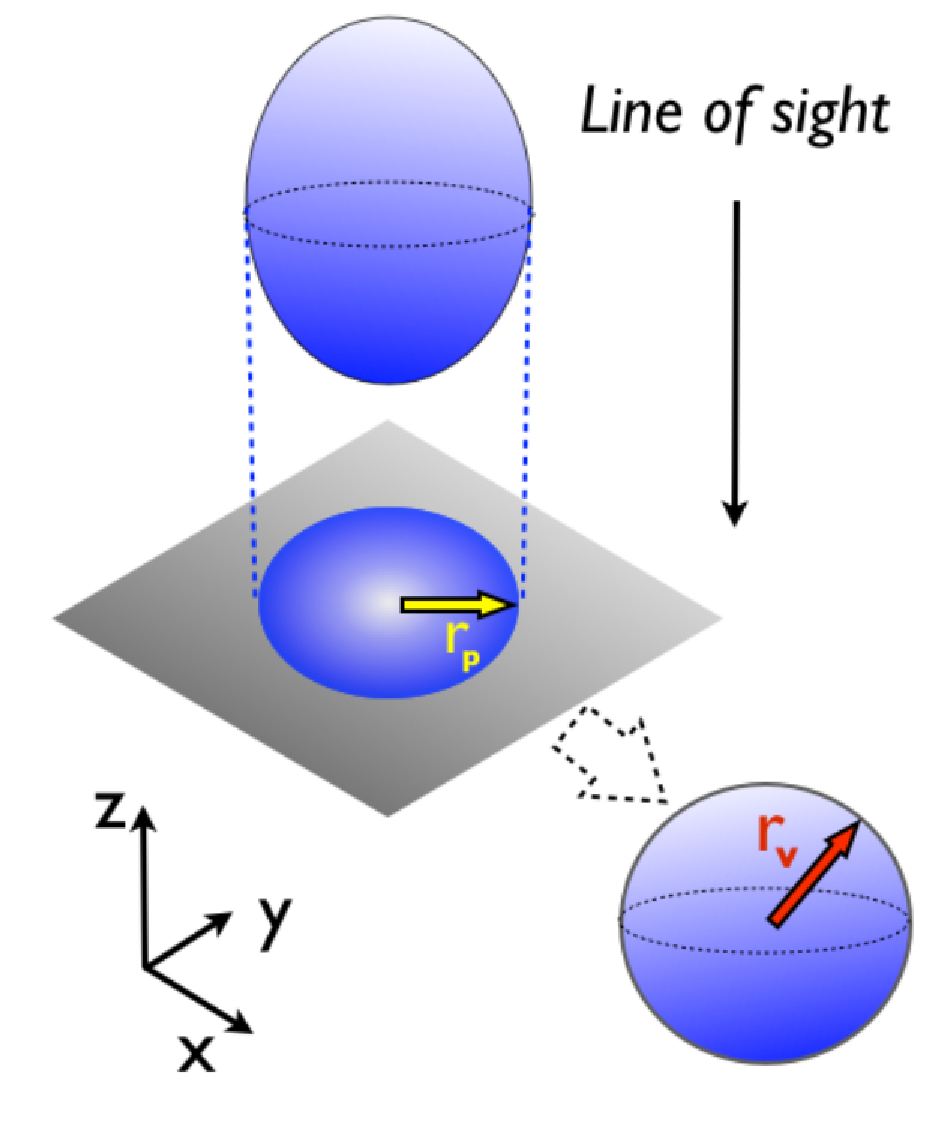} 
\caption{Representation of the method to obtain the sphere in real space from the distorted sphere in redshift space: the distorted void is projected along the line-of-sight (velocities do not affect the projection). From the projection, we reconstruct the sphere in real space. The red arrow represents $r_{\mathrm{v}}$, the radius of the void in real space; the yellow arrow $r_{\mathrm{p}}$, the radius of the projection. \label{method}}
\end{figure}

First, for a spherical void the density function is spherically symmetric. This is the density that we aim to reconstruct. We write it as $g(r_{\mathrm{v}})$, where $r_{\mathrm{v}}$ is the radius of the void, given by: $r_{\mathrm{v}}=\sqrt{x^2+y^2+z^2}$ (see Fig. \ref{method}).

Secondly, for a distorted void, the density is not spherically symmetric, since the void is distorted along the line-of-sight direction, $z$. 
For an isotropic structure, the coordinates $x$ and $y$  are invariant if we consider a rotation around the axis of the line-of-sight direction. We can then define the radius of the projection on to a plane perpendicular to the line-of-sight: $r_{\mathrm{p}}= \sqrt{x^2+y^2}$ (see Fig. \ref{method}). The distorted density is written: $\rho(r_{\mathrm{p}},z)$.

Finally we write the projected density as $I(r_{\mathrm{p}})$, only depending on the radius of the projection $r_{\mathrm{p}}$. This density can be thought as a column density. We obtain the projected density by summing galaxies in each $r_{\mathrm{p}}$ bin at all $z$ (and normalized in the bin).

We will describe in the next section the method for density profile reconstruction.

\subsubsection{The method for density profile reconstruction \label{bootsection}}
We briefly comment the steps of the method to reconstruct the density profile of the stacked void in real space (see Fig. \ref{method}). 

The first step is to project the distorted void density $\rho(r_{\mathrm{p}},z)$ along the line-of-sight in order to obtain $I(r_{\mathrm{p}})$.

The second step is to reconstruct the spherical density $g(r_{\mathrm{v}})$ from the projection $I(r_{\mathrm{p}})$. The densities $I(r_{\mathrm{p}})$ and $g(r_{\mathrm{v}})$ are related by the Abel transform, that cylindrically projects $g(r_{\mathrm{v}})$ to obtain $I(r_{\mathrm{p}})$ \citep{Abel,Bracewell}:

\begin{equation}
I(r_{\mathrm{p}})=2\int_{r_{\mathrm{p}}}^{1}\frac{g(r_{\mathrm{v}})r_{\mathrm{v}}}{\sqrt{r_{\mathrm{v}}^{2}-r_{\mathrm{p}}^{2}}}\mathrm{d}r_{\mathrm{v}}.
\end{equation}

By inverting this relation, it is possible to obtain the spherical density $g(r_{\mathrm{v}})$ from $I(r_{\mathrm{p}})$. The formula used for the reconstruction is known as the inverse Abel transform \citep{Abel,Bracewell}:

\begin{equation}
g(r_{\mathrm{v}})=-\frac{1}{\pi}\int_{r_{\mathrm{v}}}^{1}\frac{I'(r_{\mathrm{p}})}{\sqrt{r_{\mathrm{p}}^{2}-r_{\mathrm{v}}^{2}}}\mathrm{d}r_{\mathrm{p}}. \label{Abel}
\end{equation}

The problem is that the Abel inverse transform, although well mathematically defined by the formula, is strongly \textit{ill-conditioned}: if there is some noise in the input function $I(r_{\mathrm{p}})$ (of which $I'(r_{\mathrm{p}})$ is the derivative with respect to $r_{\mathrm{p}}$), the reconstruction will be dominated by noise. To overcome the problem of ill-conditioning we have implemented for the case of voids the idea proposed in \cite{Abel}, a polynomial regularization of the inversion. \cite{Durret1999} applied in the case of clusters a similar idea for the use of Abel inversion.

To check for consistency with the polynomial regularization method for the reconstruction, we also developed another method to obtain the spherical profile $g(r_{\mathrm{v}})$ using singular value decomposition. We now illustrate the two methods.

The polynomial decomposition method approximates the Abel inversion through integrals of the input function $I(r_{\mathrm{p}})$, that is directly using data.  The method allows us to manage noise in the inversion and gives good results in the case of voids, where the profile $I(r_{\mathrm{p}})$ is noisy. 

We summarize the method as follows: 
\begin{enumerate}
\item expand the spherical density to be obtained $g(r_{\mathrm{v}})$ as a polynomial series; 
\item using the polynomial expansion  of $g(r_{\mathrm{v}})$, re-write the Abel equation relating the 2D projection $I(r_{\mathrm{p}})$ and the spherical reconstruction in order to obtain a system of equations with solution $g(r_{\mathrm{v}})$;
\item solve the system of equations.
\end{enumerate}

The polynomial expansion of $g(r_{\mathrm{v}})$ is characterized by an order, $n$. The choice of the order $n$ allows us to manage noise and control the precision of the reconstruction. 
To determine the order that gives the best reconstruction we use the reprojection of the reconstructed profile: we consider the order that minimizes the difference between the $I_{\mathrm{exact}}(r_{\mathrm{p}})$ from which we reconstruct and the $I_{\mathrm{reprojected}}(r_{\mathrm{p}})$ from the reconstruction. For the application of the algorithm to real data, this test will also be possible: as we will discuss, the $I_{\mathrm{exact}}(r_{\mathrm{p}})$  is the projected density from data. Generally, for increasing $n$ the precision of the reconstruction increases and the only limitations are numerical \citep{Li2007}.

In order to avoid over or under fitting, we implement a bootstrap analysis to choose the order. Bootstrap analysis is more appropriate in a case where noise strongly affects data (as suggested by \cite{Andrae2010}). For each profile we create bootstrap samples from the sample to reconstruct. We implement the reconstruction and choose the order that gives the best fit for each one of the samples. We then take the model chosen by the different bootstrap samples. Also, to test if the choice of the order is robust, we exclude one point at a time in the profile to reconstruct and check if the chosen order is stable when redoing the analysis. Finally we also calculate the AICc information criteria \citep{Akaike1974,Akaike2002} to test the order. For the analysis of voids, the bootstrap method remains the most adapted to choose the order: it accounts for all the sources of errors such as the ill-conditioning of the inversion procedure and the errors present in the data.

The method of \cite{Li2007} assumes the boundary condition $I(1)=0$ and is described for values of the radius between 0 and 1. This is the case of the test function for the toy model, but is not the case of voids: the density is not zero outside the void. We had to adapt the method for voids by rescaling the void and considering that, if $I(r_{\mathrm{v}})$ is different from 0 in $r_{\mathrm{v}}=1$, the mean density must be subtracted from the reconstruction. Also the method described in \cite{Li2007} worked for the projection of a circular profile on a line, i.e. from 2D to 1D. We adapted it for our application of a sphere (3D) to be reconstructed from a disk (2D).

To validate the polynomial reconstruction method we control that $I(r_{\mathrm{p}})$ and $g(r_{\mathrm{v}})$ have the same value at the edge of the void, where the projection is equal to the value of the 3D function (since the projection is done along a line tangent to the void, it considers only the point at the very edge of the void).  
As a cross check for the reconstruction of the void we reproject the spherical reconstructed profile. The reprojection must match the projection of distorted density profile. 

We now illustrate the second method for the reconstruction, using the singular value decomposition approach to overcome the ill-conditioning of the Abel inverse. The singular value decomposition relies on the consideration that, if we discretize the integration of the inverse, projecting is like computing a matrix operation. We call \textbf{M} the matrix of the projection. We can write:
\begin{equation}
I=\textbf{M}G
\end{equation}
where $I$ is the projected density (that is our data, with noise), $G$ is the spherical density and \textbf{M} is the matrix allowing for the transformation between $I$ and $G$. We use singular value decomposition to decompose \textbf{M} into \textbf{U} (a unitary matrix), \textbf{W} (a diagonal matrix) and \textbf{V} (a unitary matrix). The Abel inverse can then be written as
\begin{equation}
G= \textbf{V} \textbf{W}^{-1}\textbf{U}^{T} I.
\end{equation}

The use of singular value decomposition allows us to drop the noisiest singular values, which are the smallest in matrix \textbf{W}. The number of singular values that we keep must be discussed: we need to drop enough to control noise, but not too much or we will lose information. 

The way we manage the choice of the number of dropped singular values is the same as the way we used to choose the order in the polynomial regularization method: we reproject the reconstructed profile and consider the order that minimizes the difference between the $I_{\mathrm{exact}}(r_{\mathrm{p}})$ from which we reconstruct and the $I_{\mathrm{reprojected}}(r_{\mathrm{p}})$ from the reconstruction. We use the calculation of AICc to determine the number of dropped singular values for the reconstruction. In a certain way the singular value decomposition method is the generalization of the first method without the assumption of the polynomial form for the spherical density profile to reconstruct $g(r_{\mathrm{v}})$. 
\begin{figure}
\centering
  \vspace{-5pt}
\epsfig{file=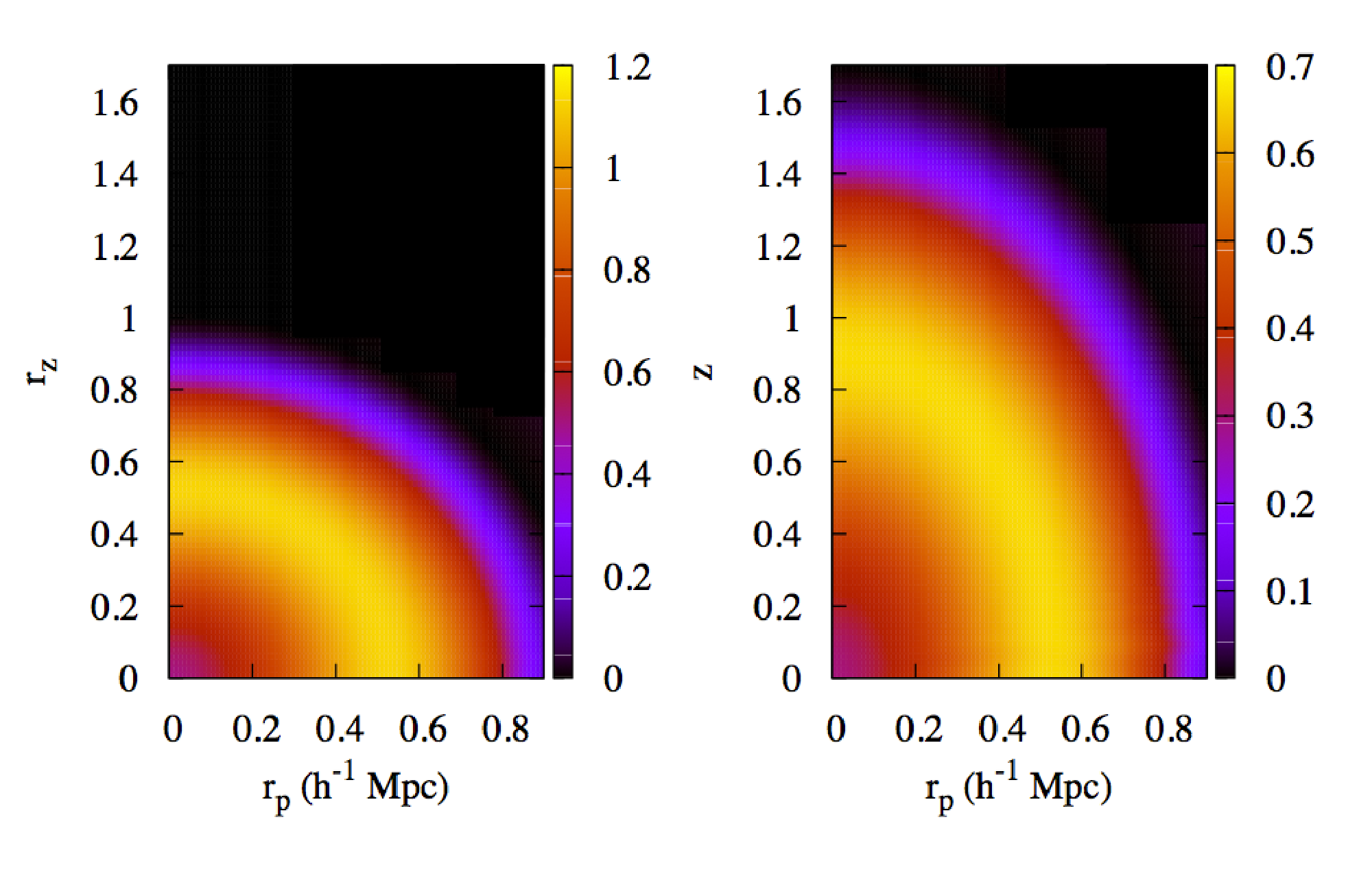,width=1\linewidth,clip=,angle=0} 
\caption{3D density spherical profile (left) and 3D density simulated distorted profile (right) for the test function. Units for the density are arbitrary in the toy model, since we use a test function. \label{step2}}
\end{figure}

There is a conceptual difference between the two methods. The singular value decomposition method determines the basis that gives the best reconstruction using all the points of $I(r_{\mathrm{p}})$ to calculate the spherical density. Thus it gives a more regular reconstructed density profile for the first points.  The determination is however strongly dependent on data and might be more sensitive to noise. On the other hand, the method with polynomial regularization of the Abel inverse enforces polynomial smoothness and calculates the values of the density $g(r_{\mathrm{v}})$ at each point, considering for the calculation only the points  of $I(r_{\mathrm{p}})$ from the considered radius $r_{\mathrm{p}}$ to the edge of the sphere (see \cite{Li2007} for details). A separate reconstruction for each point of $g(r_{\mathrm{v}})$ gives a less regular profile for the first points of the profile (due to the higher difficulty of disentangling the 3D structure from a projection when considering all the radii from the centre to the edge, as it is for the inner points) but might be useful to control noise for the reconstruction  of voids, where the presence of clumps in the wall and noise in data is likely to affect the quality of the reconstruction.

In the next sections, we apply the reconstruction to a toy model and a dark matter simulation.

\begin{figure}
\centering \includegraphics[width=0.3 \textwidth,angle=-90]{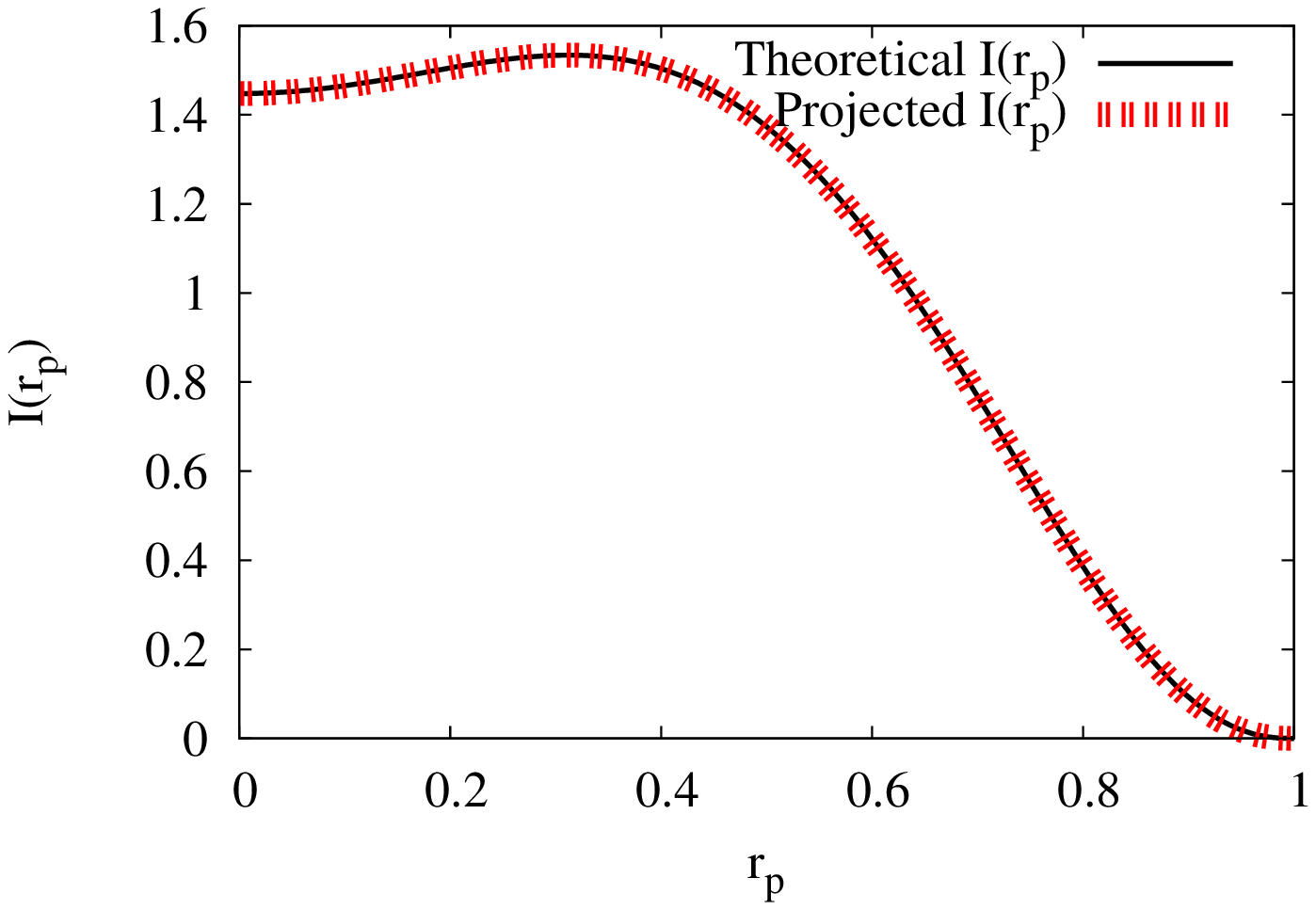}
\caption{Projection from a 3D simulated distorted profile (red bars) and theoretical projection (black line). As a sanity check: the projected profile from the distorted sphere matches the profile from the theoretical projection (the projection of the spherical profile), the projection cancels the deformation of the density profile.\label{I}}
\end{figure}

\subsection{Testing the method with toy model}
In order to test the feasibility of the method, we can simulate a distorted profile by artificially adding a velocity along the line-of-sight to a spherical profile. Since we know the initial spherical profile, we can test our algorithm by trying to recover the correct initially spherical density from the distorted one. We use the simplicity of this toy model to illustrate the full method for the reconstruction of the spherical density profile, so that in the next sections we can directly present results for simulations and real voids.

From the presentation and explanation of the method in previous sections, it can be understood that the following steps are necessary: create a distorted profile, project it along the line-of-sight and reconstruct the sphere from the projection.

 \begin{figure}
\centering \includegraphics[width=0.33 \textwidth,angle=-90]{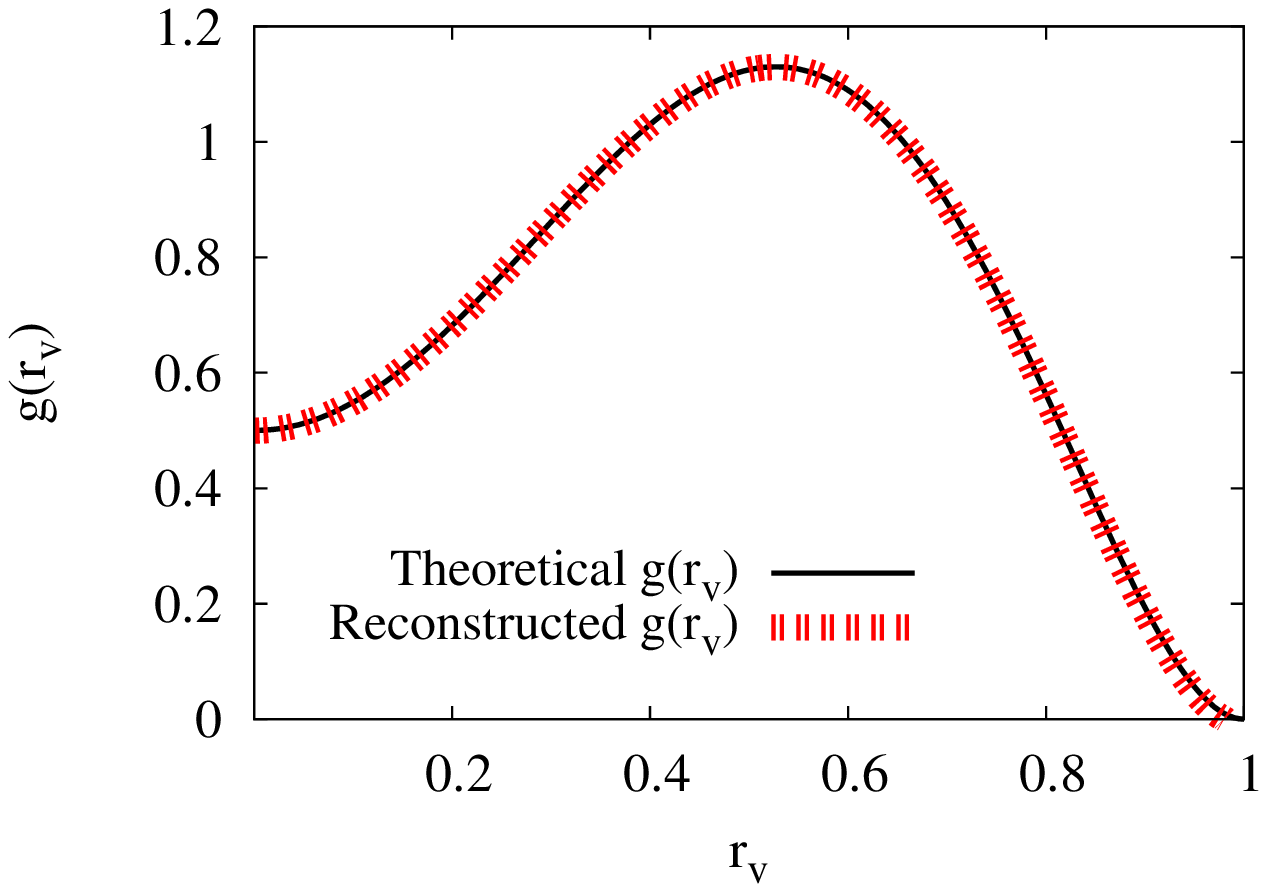}
\caption{Theoretical profile of the 3D density $g(r_{\mathrm{v}})$ (black line) and reconstructed profile (red bars) in the case without noise (using the method of polynomial regularization). \label{Nonoise}}
\end{figure}

\begin{figure}
\hspace{-2.5em}
\includegraphics[width=0.39 \textwidth,angle=-90]{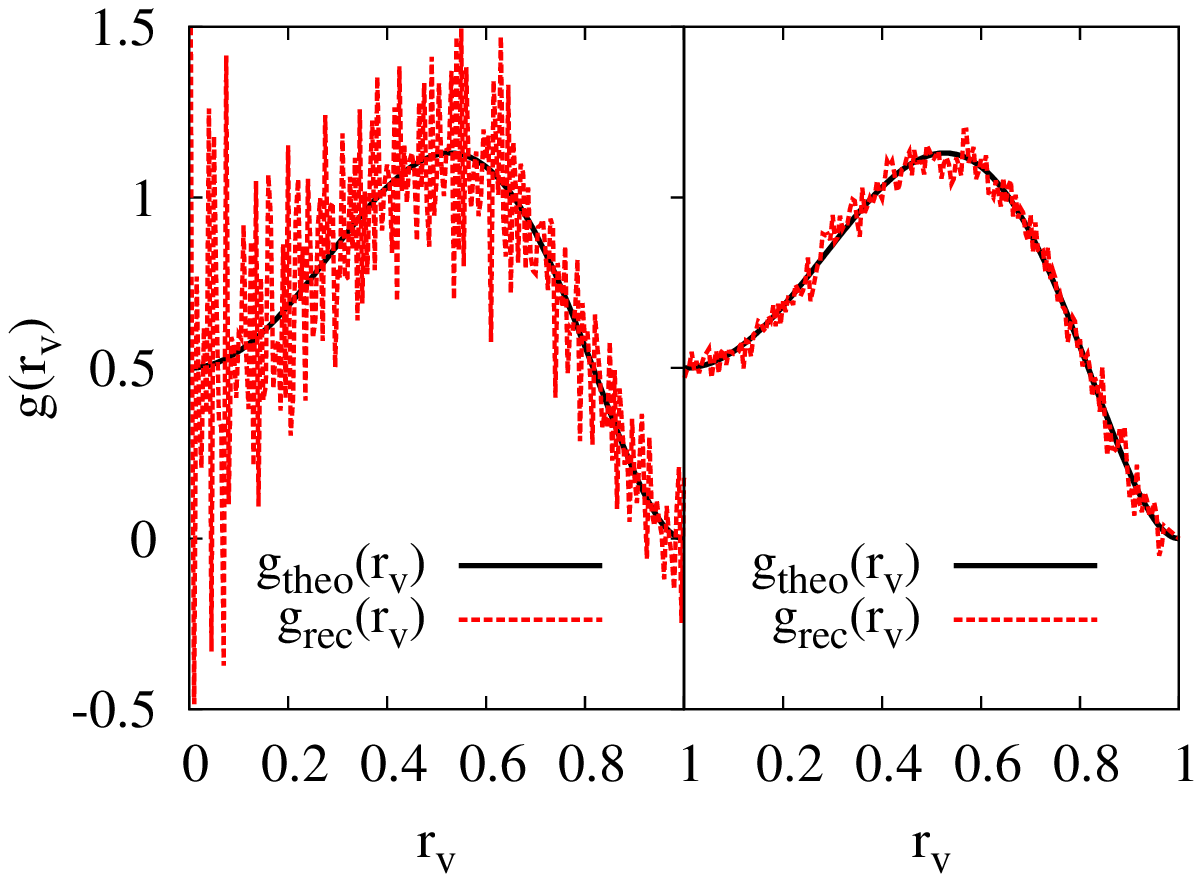}
\caption{Theoretical profile of the 3D density $g(r_{\mathrm{v}})$ (black line) and reconstructed profile (red) in the case with 1\% Gaussian noise  in the input function. The left-hand plot shows the Abel inversion without regularization, the right-hand plot shows the reconstruction obtained with the polynomial regularization of the inversion.\label{Gaunoise}}
\end{figure}

In order to have an efficient test, we choose an example function for which we can calculate the exact Abel inverse through mathematical integration. These kinds of functions are called Abel pairs \citep{Abel, Bracewell}. We test all the steps of the algorithm with this function, considering that we know through analytic calculation $g_{\mathrm{exact}}(r_{\mathrm{v}})$ and $I_{\mathrm{exact}}(r_{\mathrm{p}})$, related through Equation (\ref{Abel}). We have chosen the following test function:
\begin{eqnarray}
I_{\mathrm{exact}}(r_{\mathrm{p}})=\frac{8}{105}\sqrt{1-r_{\mathrm{p}}^2}(19+34r_{\mathrm{p}}^2-125r_{\mathrm{p}}^4+72r_{\mathrm{p}}^6)\\
g_{\mathrm{exact}}(r_{\mathrm{v}})=\frac{1}{2}(1+10r_{\mathrm{v}}^2-23r_{\mathrm{v}}^4+12r_{\mathrm{v}}^6).
\end{eqnarray}
The function for the toy model needs to have an exact mathematical inversion, this is the only important constraint for its choice. Additionally, it has a shape whose features roughly match those of a void profile.

The first step is to create a distorted profile from the spherical profile $g_{\mathrm{exact}}(r_{\mathrm{v}})$. We show the results of the distortion in Fig. \ref{step2} (right-hand plot), along with the spherical profile (left-hand plot). The void is distorted by adding an artificial velocity component to the $r_{z}$ coordinate (as described in equation \ref{vel}), which, as expected, changes the value of the density. 

\captionsetup[figure]{labelfont=it,textfont={bf}}
\renewcommand{\tabcolsep}{-1.5cm}
\begin{figure*}
\begin{tabular}{cc}

  \centering  
 \hspace{7.5em}
  \vspace{-10pt}
  \includegraphics[width=0.9\columnwidth, angle=-90]{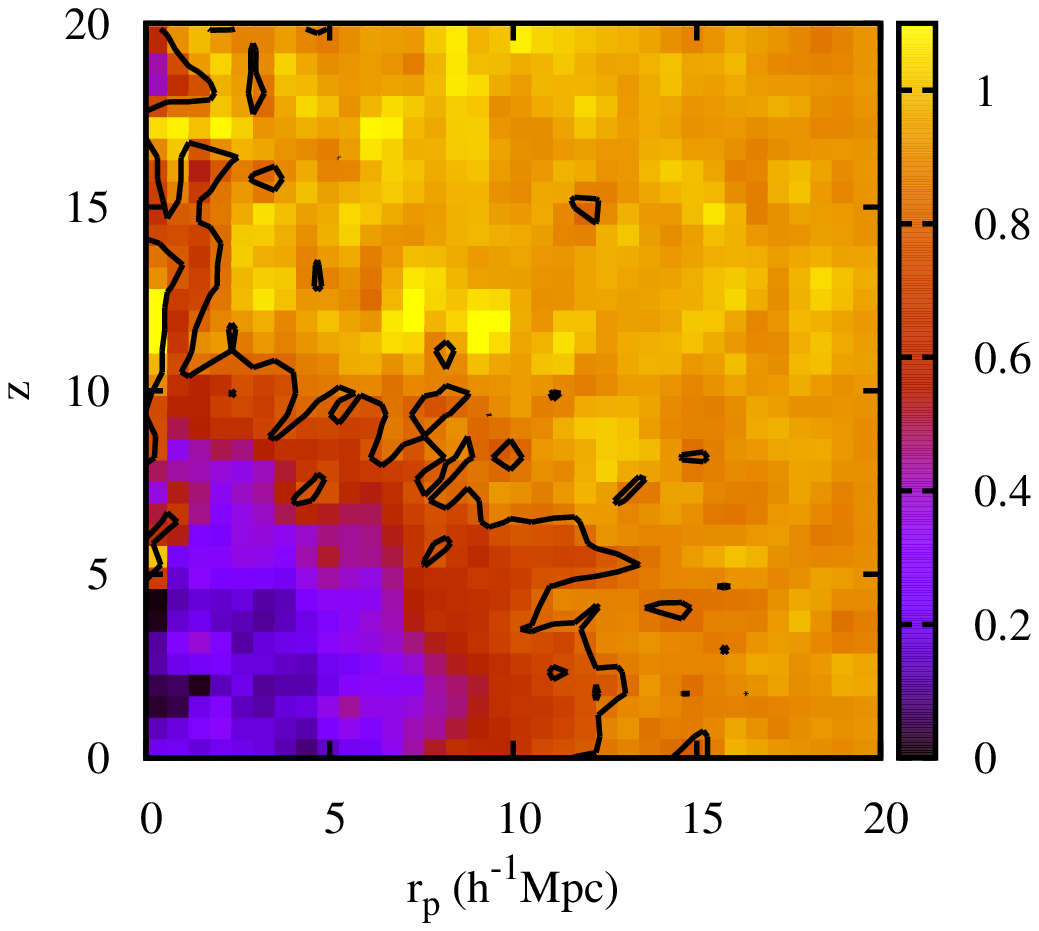}&
    \includegraphics[width=0.9\columnwidth, angle=-90]{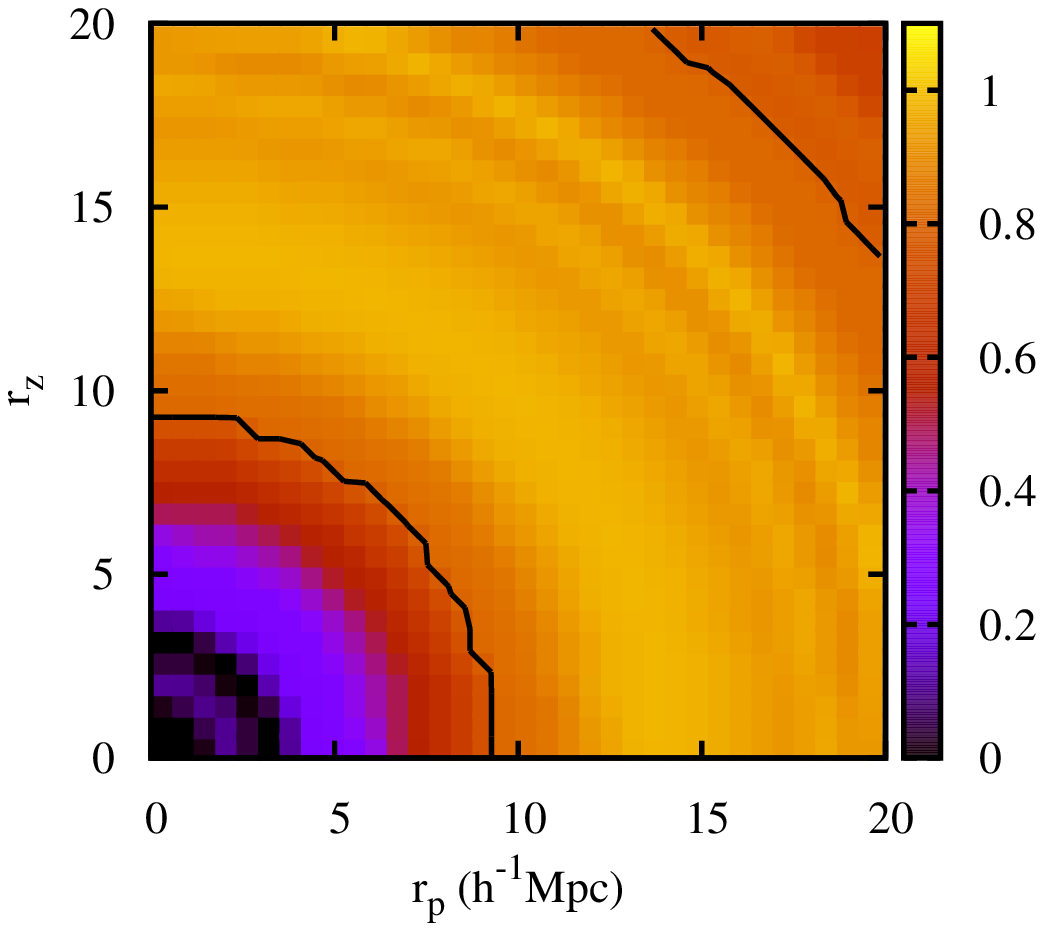}\\
  
  \label{fig:example}
  \end{tabular}
 \caption{Distorted density profile of stacked void (left) from simulation and reconstructed spherical void in real space (right), both normalized to the mean density. Black contours in both images are density contours at 0.8 (where we have normalized to mean density). }
 \label{Rho_sim}
 
\end{figure*}

The next step is the projection of the distorted profile. Peculiar velocities contribute to redshift and distort the density profile; but, since the distortion is along the line-of-sight, velocities do not affect the projection.
As a sanity check, we control that the projection of the distorted density is the same as the projection $I_{\mathrm{exact}}(r_{\mathrm{p}})$ from the non-distorted profile $g_{\mathrm{exact}}(r_{\mathrm{v}})$, even when using different kinds of velocity to distort the profile (such as $v(r_{\mathrm{v}})=ar_{\mathrm{v}}$, or $v(r_{\mathrm{v}})=ar_{\mathrm{v}}^2$). Fig. \ref{I} shows the result of the comparison: the profiles match perfectly.

Once we have the projection, we can reconstruct the spherical density profile of the stacked void, $g(r_{\mathrm{v}})$. 

We show in Fig. \ref{Nonoise} an example of the reconstruction of $g(r_{\mathrm{v}})$ from the test function $I(r_{\mathrm{p}})$ without noise. To show the ability of the algorithm to reduce noise in the reconstruction, we show the reconstruction in the case of a 1\% Gaussian noise in the input function and compare this to the direct calculation of Abel inverse, without methods to reduce the noise (see Fig. \ref{Gaunoise}). The reconstruction with regularization matches the theoretical $g_{\mathrm{exact}}(r_{\mathrm{v}})$.\\

In this simple case, because the function can be inverted analytically, both the singular value decomposition and the polynomial reconstruction method give very good results (the reconstruction overlaps with the theoretical profile). We widely tested the spherical reconstruction with the methods for many known functions (not only our test function), both without noise and with noise (we added a 1\%, 3\% and 5\% noise to other test functions and correctly reconstructed the 3D profile). In the next section, we discuss the presence of noise in the profile and argue that a full dark matter simulation is needed to correctly test the reconstruction algorithm.

\subsection{Noise in density profiles}

In the case of the toy model, we have considered an arbitrary percentage of noise, aiming to assess the capability of the algorithm to overcome noise in the reconstruction. 

 Despite of its capacity to show noise reduction in the inverse, the toy model cannot account in a realistic and physical way for the complex sources of noise that would be present in a full simulation. 
The main source of noise in the density profiles is due to the sparsity of data, specifically Poisson noise on galaxy counts in the bins for the projected $I(r_{\mathrm{p}})$. The use of the stacking procedure allows us to obtain well-populated stacks, thereby permitting the extraction of cosmological information.

So, while the simpler case of the toy model is a proof of concept to assess the capability of the algorithm to control noise in the reconstruction procedure, the use of a simulated stacked void accounts for a more complex and realistic  situation, where noise is implicitly taken into account. Furthermore, the use of a simulated void from a full dark matter particle simulation naturally takes into account the clustering of structures, serving the purpose of this paper to test the reconstruction algorithm and show its first application as a proof of concept. 

The simulation provides us with a robust test for the reconstruction algorithm and for the impact of noise in the reconstruction. More details are given in the next section.

\begin{figure}
\centering
\epsfig{file=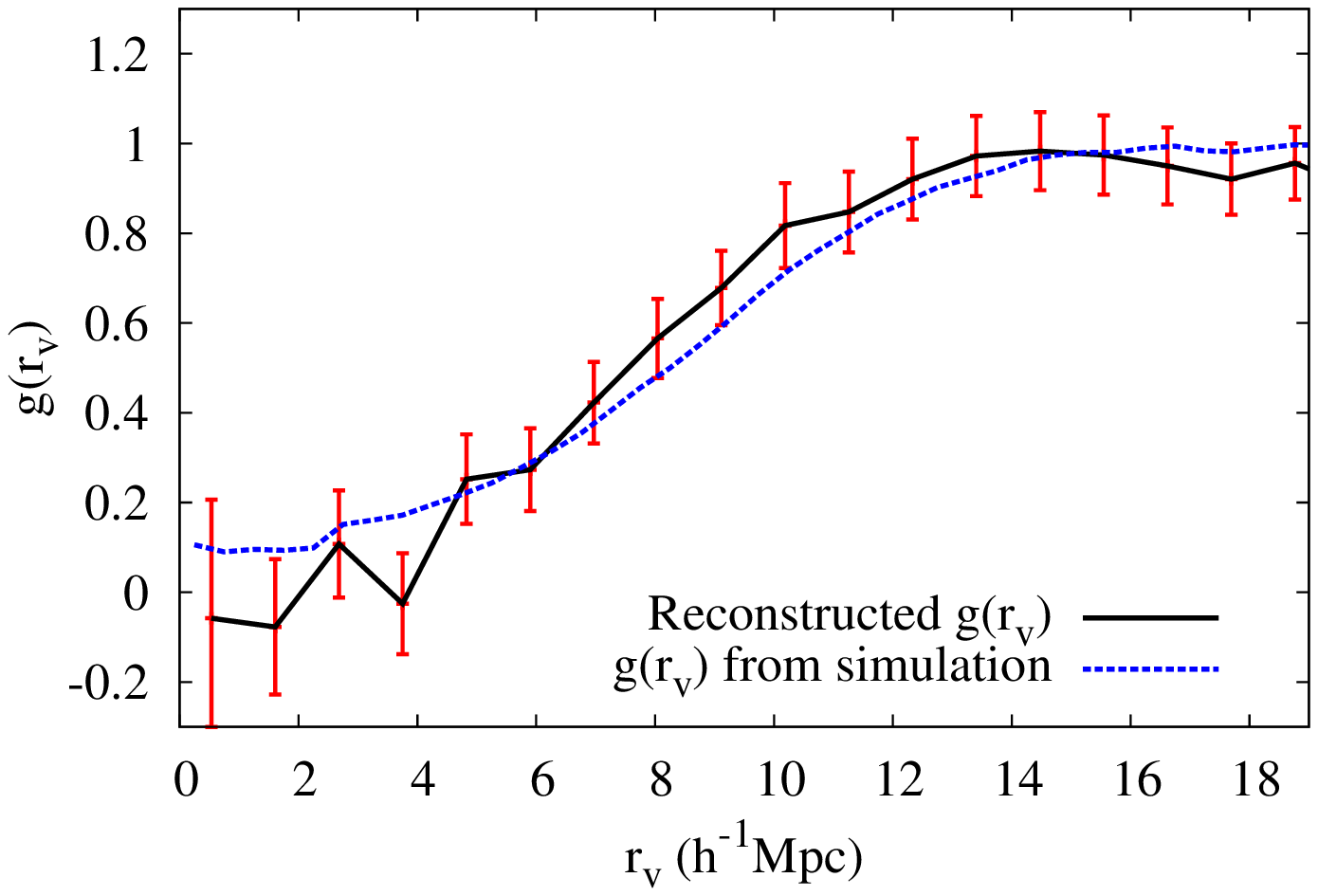,width=0.7\linewidth,clip=, angle=-90} 
\caption{The polynomial reconstruction matches the spherical profile from simulation within the error bars (except for the inner part of the profile, as discussed in Section \ref{sectionsim}). A further confirmation of the agreement is given by the match of the projection (see Fig. \ref{riproiezione}). The reconstruction is obtained from a subsample of 200,000 dark matter particles of the total (about $10^{9}$ particles). The error bars are correlated. \label{compare}}
\end{figure}

\begin{figure}
\centering
\epsfig{file=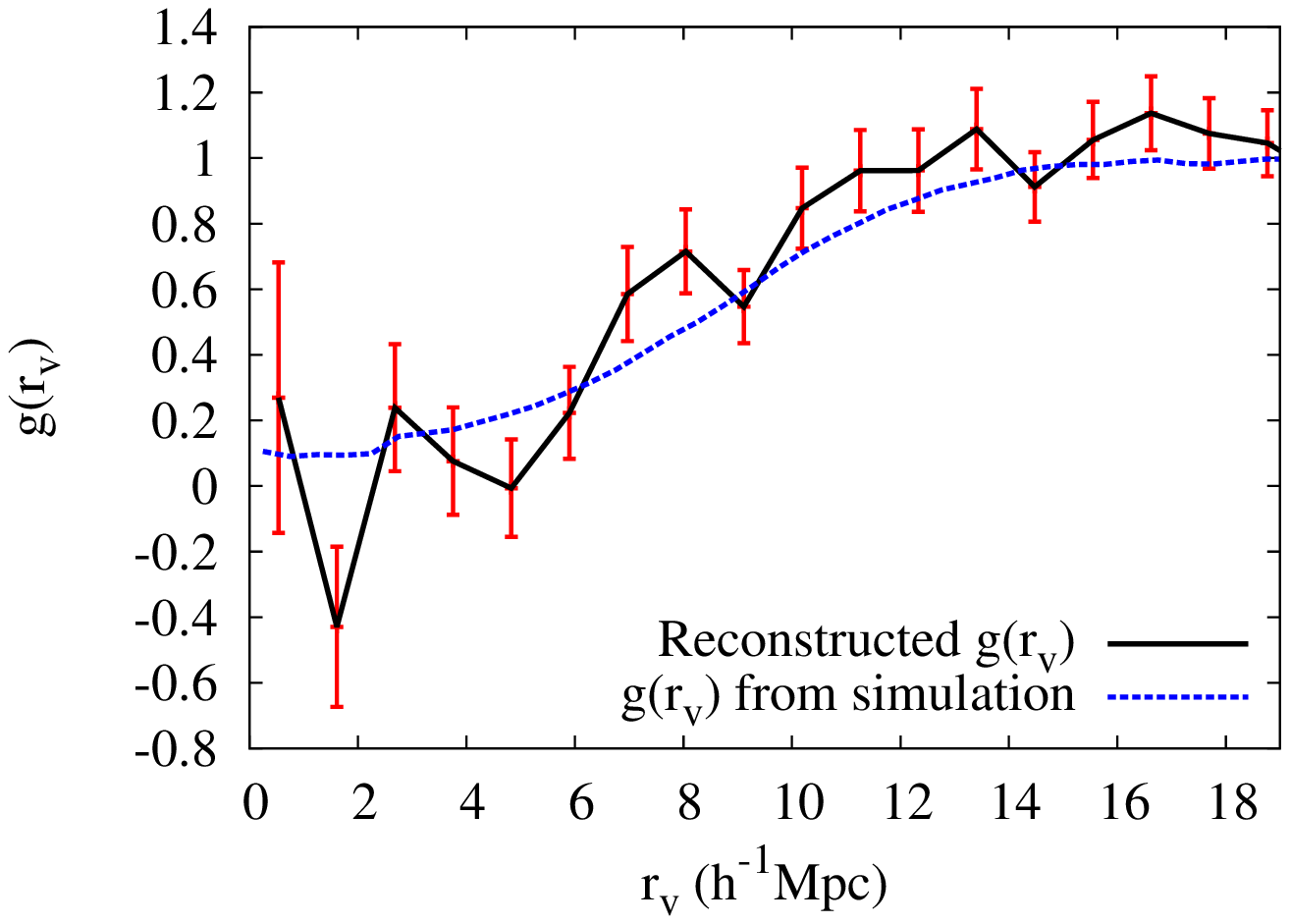,width=0.7\linewidth,clip=, angle=-90} 
\caption{Reconstructed density for the simulated void from a smaller subsample (100,000 dark matter particles of the total, about $10^{9}$ particles).\label{riproiezione1}}
\end{figure}

\begin{figure}
\centering
\epsfig{file=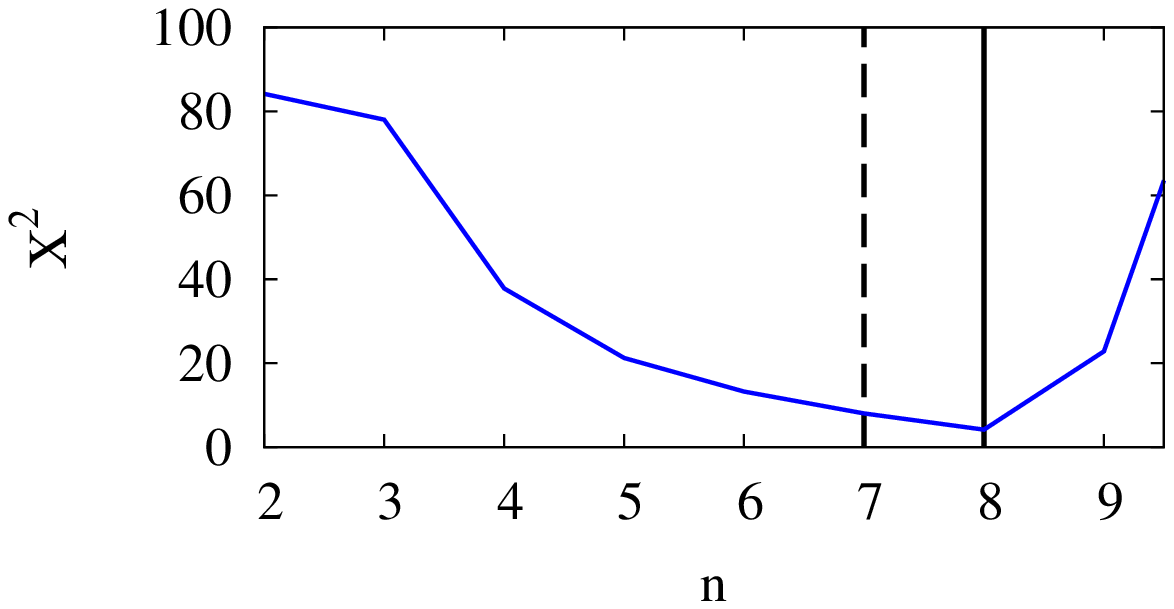,width=0.7\linewidth,clip=, angle=-90} 
\caption{Choice of the order for the polynomial regularization method of the Abel inverse, in the case of a simulated void. The solid black line is the order chosen by bootstrap method, which also coincides with the order chosen by the AICc information criterion and minimization of chi-squared. The dashed black line shows the order chosen by minimizing the reduced chi-squared. \label{chifig}}
\end{figure}

 \section{Testing the method with a simulated stacked void}\label{sectionsim}
 
We will now compare the reconstruction methods in a more realistic case: a stacked void from a full dark matter simulation. 
We test the reconstruction in the case of a full simulation (by comparison with the known spherical profile from the simulation) and we show the consistency between results from the two reconstruction methods. 

The simulated stacked void contains voids with radii between 10 and 12 $\mathrm{h^{-1}Mpc}$ from a dark matter particle simulation in a 500 $\mathrm{h^{-1}Mpc}$ box with $512^{3}$ particles used in \cite{Lavaux2012}. The void finder is also the same, based on \cite{Neyrinck2008} (\tt ZOBOV\normalfont). We clearly see the void profile (Fig. \ref{Rho_sim}, left-hand plot) in redshift space, with a low density at the centre and a wall at 10-12 $\mathrm{h^{-1}Mpc}$. As expected, the distortion is along the line-of-sight direction.

\subsection{Reconstructed density profile of simulated stacked void}

The spherical reconstructed profile is shown in Fig. \ref{Rho_sim} (right). To test the quality of the reconstruction we use the known spherical profile from the real-space position of the particles. Fig. \ref{compare} shows the result of the reconstruction: it matches the spherical profile from simulation, validating the reconstruction. It must be noted that the reconstruction is obtained from a subsample of 200,000 dark matter particles of the total (about $10^{9}$ particles). Real stacked voids do not have $10^{9}$ galaxies as the simulated stacked void and, by taking only 200,000 of $10^{9}$, we crudely simulate the effect of sub-sampling due to the fact that we are not able to observe all the galaxies that shape voids. We also show in Fig. \ref{riproiezione1} a reconstructed profile obtained from a sample of 100,000 particles in the same void stack: the reconstruction is noisier and with higher errors, but we are still able to reconstruct the void shape despite the smaller subsampling. This shows the capability of the algorithm to work with a subsampled number of galaxies, as in the case of real stacked voids. Furthermore the quality of the reconstruction can be assessed by checking the reprojection of the profile.

We compute error bars for the polynomial reconstruction method considering Poisson noise on galaxy counts in the bins for the projected $I(r_{\mathrm{p}})$ and use the bootstrap method to obtain the error bars in the reconstruction and in the reprojection. The bootstrap error analysis gives a realistic estimation of errors due to the finite number of galaxies. We show in Fig. \ref{chifig} the choice of the order for the simulated void reconstruction (following the procedure discussed in Section \ref{bootsection}). The order selected by the bootstrap method is the most realistic to choose, since the bootstrap analysis takes into account all the errors affecting the reconstruction. 

The estimates for the density profile reconstruction are correlated. The error bars are higher at small radii of the void because the algorithm of polynomial regularization is less precise for inner points: the reconstruction is more complicated at the centre, where the projection gets a major contribution from the outer shells of the sphere. 

Before concluding this section, we briefly comment the differences between the toy model and the simulation reconstructions. In the toy model the simplicity of the function used to roughly represent a density profile of a void gives rise to regular contours even after the distortions due to the added peculiar velocities. The contours in Fig. \ref{step2} remain symmetric. On the contrary, the simulated stacked void has all the complexity of a real stacked void, including realistic noise in the projected shape of the void that we use to reconstruct the spherical density profile in real space. The presence of noise results in contours that have a slightly different extent in $r_{\mathrm{p}}$ compared to the corresponding redshift profile (see Fig. \ref{Rho_sim}). 

Despite the presence of this kind of effect, arising in the realistic case of the simulation, the reconstruction algorithm still dominates the ill-conditioning of the inverse and is able to manage noise, obtaining a profile that is coherent (as discussed in this and in the next section) with the profile from the simulation, used to test the reconstruction.

\begin{figure}
\centering
\epsfig{file=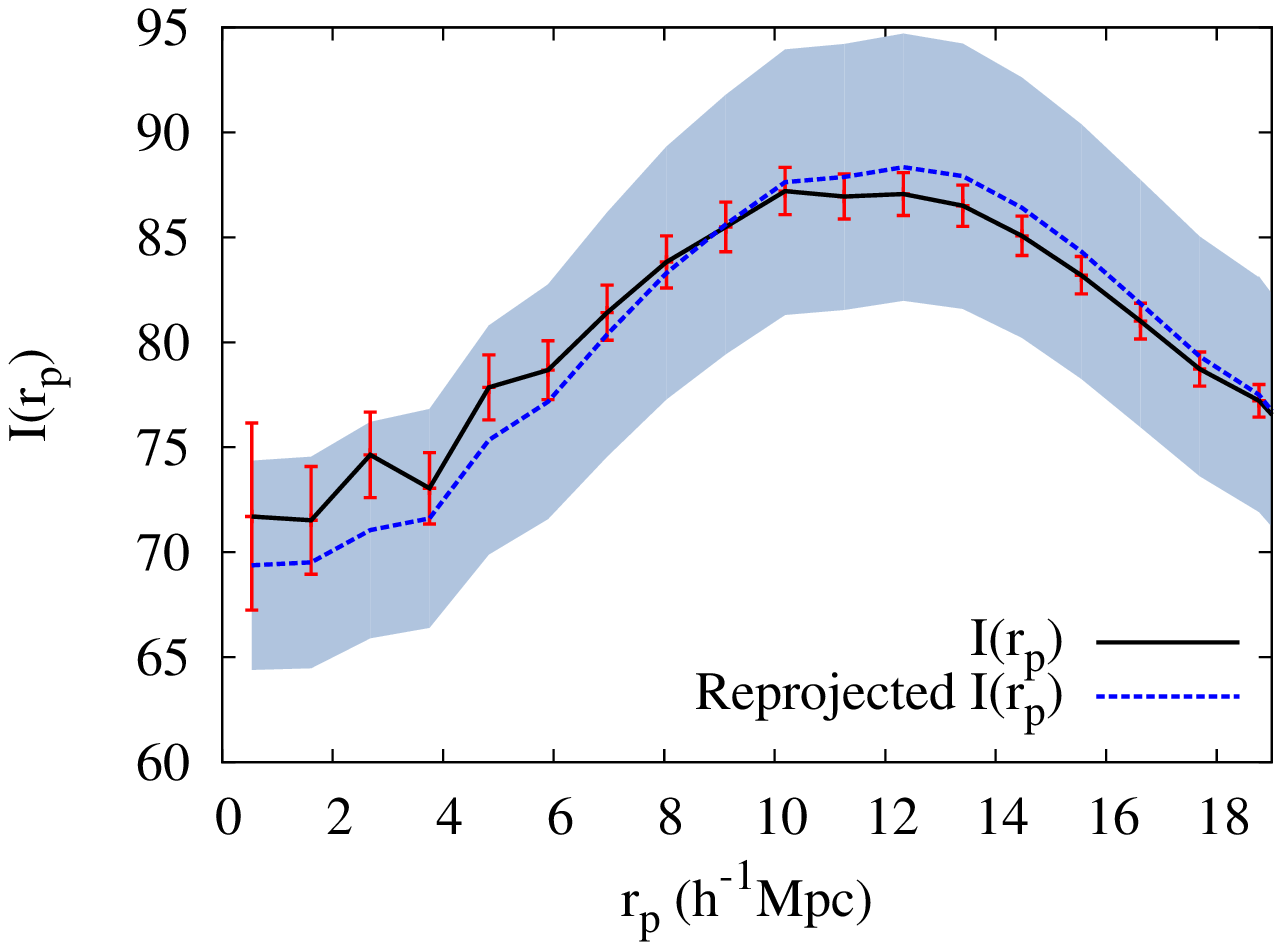,width=0.7\linewidth,clip=, angle=-90}  
\caption{For the simulated void, match between the $I(r_{\mathrm{p}})$ from simulated data and the reprojection from the reconstructed profile from a subsample of 200,000 dark matter particles of the total (about $10^{9}$ particles). The light-blue bands are the errors on the reprojected $I(r_{\mathrm{p}})$ (that is obtained by projecting the reconstructed spherical density profile $g(r_{\mathrm{v}})$). \label{riproiezione}}
\end{figure}

\begin{figure}
\centering
\epsfig{file=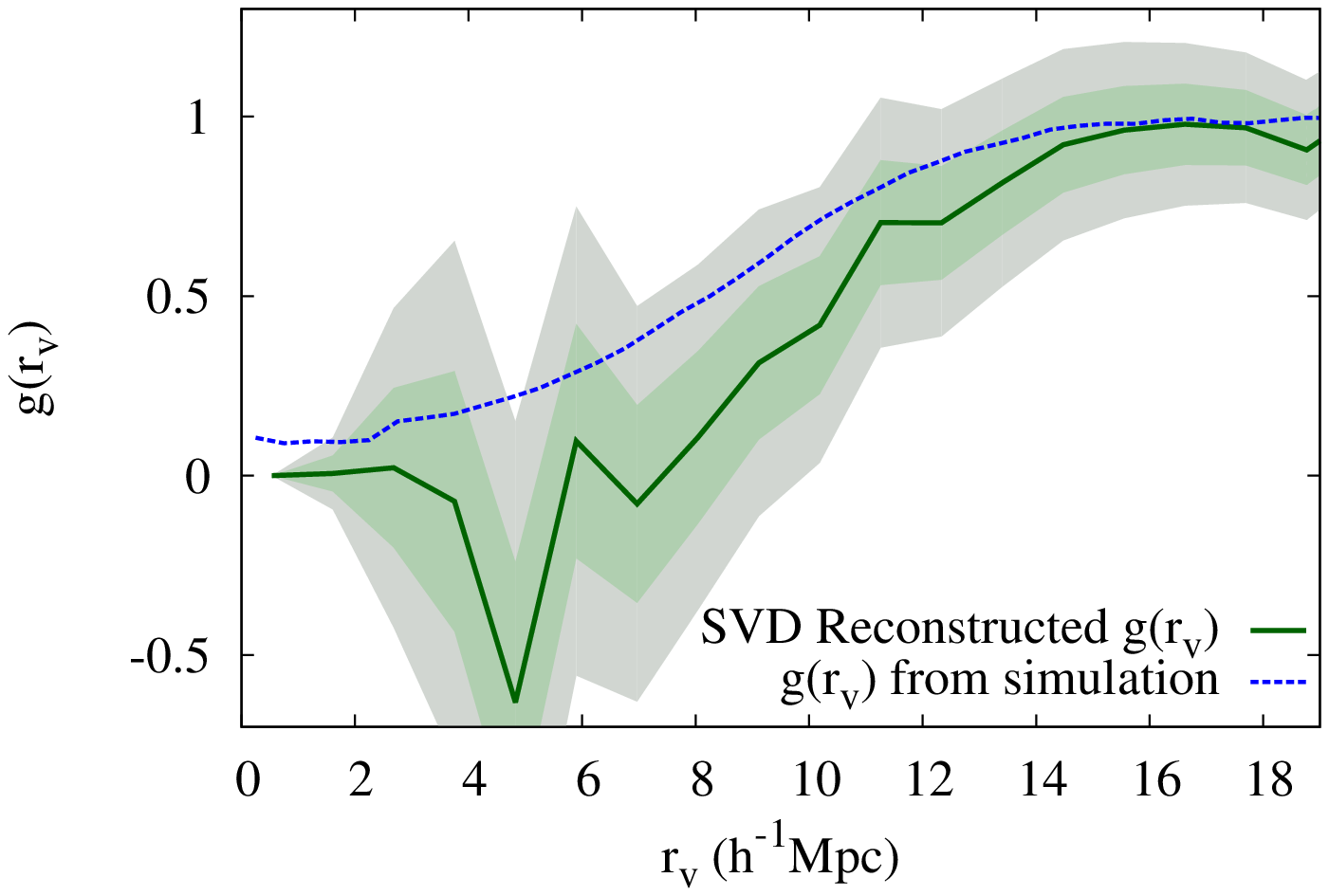,width=0.7\linewidth,clip=, angle=-90} 
\caption{The singular value decomposition reconstruction matches the spherical profile from simulation within the error bars (green bands correspond to $1\sigma$, grey to $2\sigma$), but is more affected by noise than the polynomial regularization method. The reconstruction is obtained from a subsample of 200,000 dark matter particles of the total (about $10^{9}$ particles). The error bars are correlated.\label{compare_SVD}}
\end{figure}

\subsection{The reprojection, a quality test for the reconstruction}\label{sanity}\label{reprojection}

To further check for consistency we also reproject the reconstructed spherical void (Fig. \ref{riproiezione}). This is an important sanity check for the reconstruction algorithm. In ill-conditioned problems, noise can easily blow up and completely dominate the results. 

For this particular problem of reconstruction, we have the possibility to re-invert the procedure by projecting the reconstructed density profile to check if its projection matches the projected profile $I(r_{\mathrm{p}})$  from which we made the reconstruction. In the case of data with noise, the consistency test allows us to check results: the match of the reprojection can be used to validate the reconstruction for the profiles when applying the algorithm to real data, where the ill-conditioning due to noise must be dominated. 
So, in addition to the robustness of the method (that uses chi-square, AICc criteria and also bootstrap analysis to obtain a profile acceptable within the error bars), we have here an independent quality test validating the reconstruction.

Fig. \ref{riproiezione} shows the result of this test for the simulated void: the reprojection matches the initial projection $I(r_{\mathrm{p}})$ (within the error bars), validating the reconstruction. The $I(r_{\mathrm{p}})$ is obtained from the simulation, by projecting the positions of galaxies and counting galaxies in radial bins on the plane of the projection. While the inner points of the profile are noisier as expected, we get high-quality information for the part of the void where the density rises from low to high values near the wall. \\

\subsection{The singular value decomposition method for the simulated void}

We also show in Fig. \ref{compare_SVD} the reconstruction with the singular value decomposition method, in order to check for consistency. 
As discussed, the profile obtained in the case of the singular value decomposition method is more sensitive to the presence of clumps in the wall, because it considers all the points together to obtain the profile $g(r_{\mathrm{v}})$. This might affect the quality of reconstruction. Furthermore, the singular value decomposition method has larger error bars since it does not use prior information (except the truncation of the matrix of singular values); while the polynomial regularization method enforced polynomial smoothness. For this reason the singular value method is less precise than the polynomial method.

As a conclusion, apart from the mentioned difference, both methods (polynomial regularization and singular value decomposition) allow us to manage noise in the Abel inverse transform and show similar reconstructed profiles. For practical purposes we have chosen the polynomial regularization method, that is more adapted in the case of voids, and use the second to check for consistency in the reconstruction.

The reconstruction of the spherical profile for stacked voids in the case of a dark matter particle simulation  (Fig. \ref{riproiezione}) is completely implemented and tested. As a further test of the quality of the reconstruction and capability of the algorithm, we describe in the next section a test with stacked voids from a mock galaxy catalogue.\\
\renewcommand{\tabcolsep}{-1cm}

\section{Testing the algorithm with stacked voids from a mock galaxy catalogue}

To further test the capability of the reconstruction algorithm, we use a mock galaxy catalogue matching the properties of the SDSS DR7. The mock catalogue is sourced from a high -resolution $N$-body dark matter simulation with $\Lambda$CDM cosmology, $1024^{3}$ particles and 1 $h^{-1} \mathrm{Gpc}$ side (also used in \cite{Sutter2013}) and part of the Public Cosmic Void Catalog\footnote{http://www.cosmicvoids.net}. The cosmological parameters of the simulation assume a WMAP 7-year cosmology, the initial conditions of the simulation were obtained through a power spectrum calculated with \tt{CLASS} \normalfont\citep{Blas2011} and realized with a modified version of \tt2LPTIC \normalfont\citep{Crocce2006}. The simulation is used as a source for a Halo Occupation Distribution model \citep{Tinker2006, Zheng2007} to produce the galaxy catalogue. The model assigns to each dark matter halo of mass $M$ a central galaxy and satellite galaxies, the mean number of central galaxies and satellites is described by:

\begin{eqnarray}
\big \langle N_{\rm{cen}}(M)\big \rangle =\frac{1}{2}\bigg[ 1+\rm{erf}\bigg(\frac{\rm{log} \it M-\rm{log} \it M_{\mathrm{min}}}{\sigma_{\rm{log} \it M}}\bigg)\bigg ]\\
\big \langle N_{\rm{sat}}(M)\big \rangle =\big \langle N_{\rm{cen}}(M)\big \rangle \bigg(\frac{M-M_{0}}{M^{'}_{1}}\bigg)^{\alpha},
\end{eqnarray}
where we have $\sigma_{\rm{log} \it M}$, $M_{\rm min}$, $M_{0}$,$M_{1}^{'}$ and $\alpha$ as free parameters which are set to match the properties of a given galaxy population. Namely, we match the galaxy population to the main sample of SDSS DR7 \citep{Strauss2002, Zehavi2011}. 

This allows us to have a mock galaxy catalogue exactly matching the real data to which we will apply the reconstruction algorithm. We thus run the void finder \tt VIDE \normalfont described in \cite{Sutter2014a} and obtain void stacks on which we run the reconstruction with polynomial regularization. 

With the methodology described in the previous section, we apply the algorithm to stacked voids obtained from the mock galaxy catalogue matching the properties of the SDSS DR7. To assess the capability of the algorithm, we compare the reconstructed profile with the real-space profile of the stacked void from the mock catalogue. Furthermore, we use the reprojection of the profile as a quality test for the reconstruction, as described in Section \ref{reprojection}. This independent test is a further validation of the reconstruction.

\begin{figure}
\centering
\epsfig{file=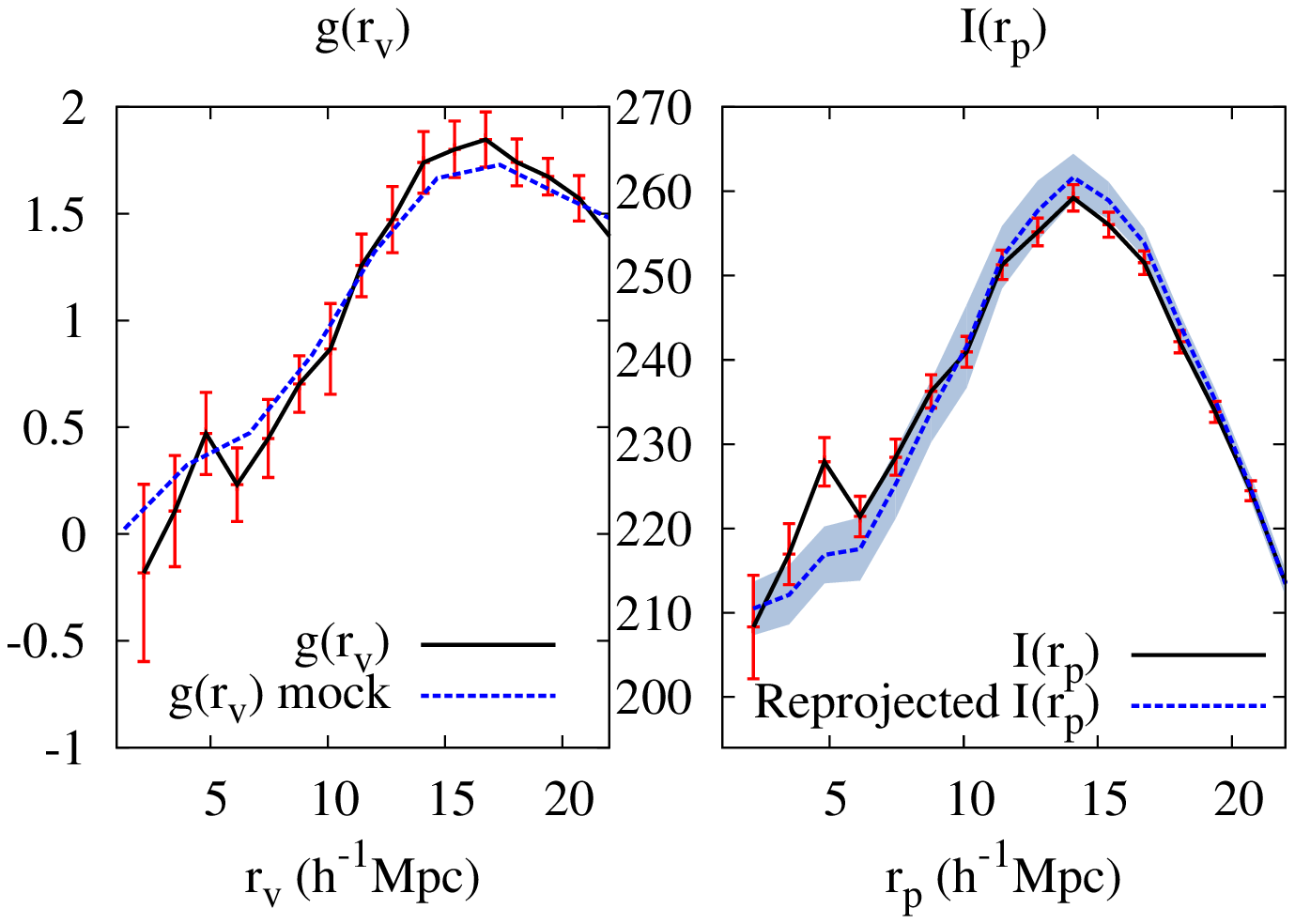,width=0.7\linewidth,clip=, angle=-90}  
\caption{Reconstruction for a 10-15 $\mathrm{h^{-1}Mpc}$ stacked void from the mock galaxy catalogue: left-hand plot shows the match between the profile in real space from the mock catalogue (dashed blue line) and the reconstructed profile $g(r_{\mathrm{v}})$(black line); right-hand plot shows the match between the $I(r_{\mathrm{p}})$ from the mock catalogue (black line) and the reprojection from the reconstructed profile $g(r_{\mathrm{v}})$ (dashed blue line). The light-blue bands are the errors on the reprojected $I(r_{\mathrm{p}})$ (that is obtained by projecting the reconstructed spherical density profile $g(r_{\mathrm{v}})$). Here, we have normalized to mean density for $g$ (while $I(r_{\mathrm{p}})$ units are number of galaxies per  $\mathrm{(h^{-1}Mpc)^{2}}$).  \label{hod10-15}}
\end{figure}

\begin{figure}
\centering
\epsfig{file=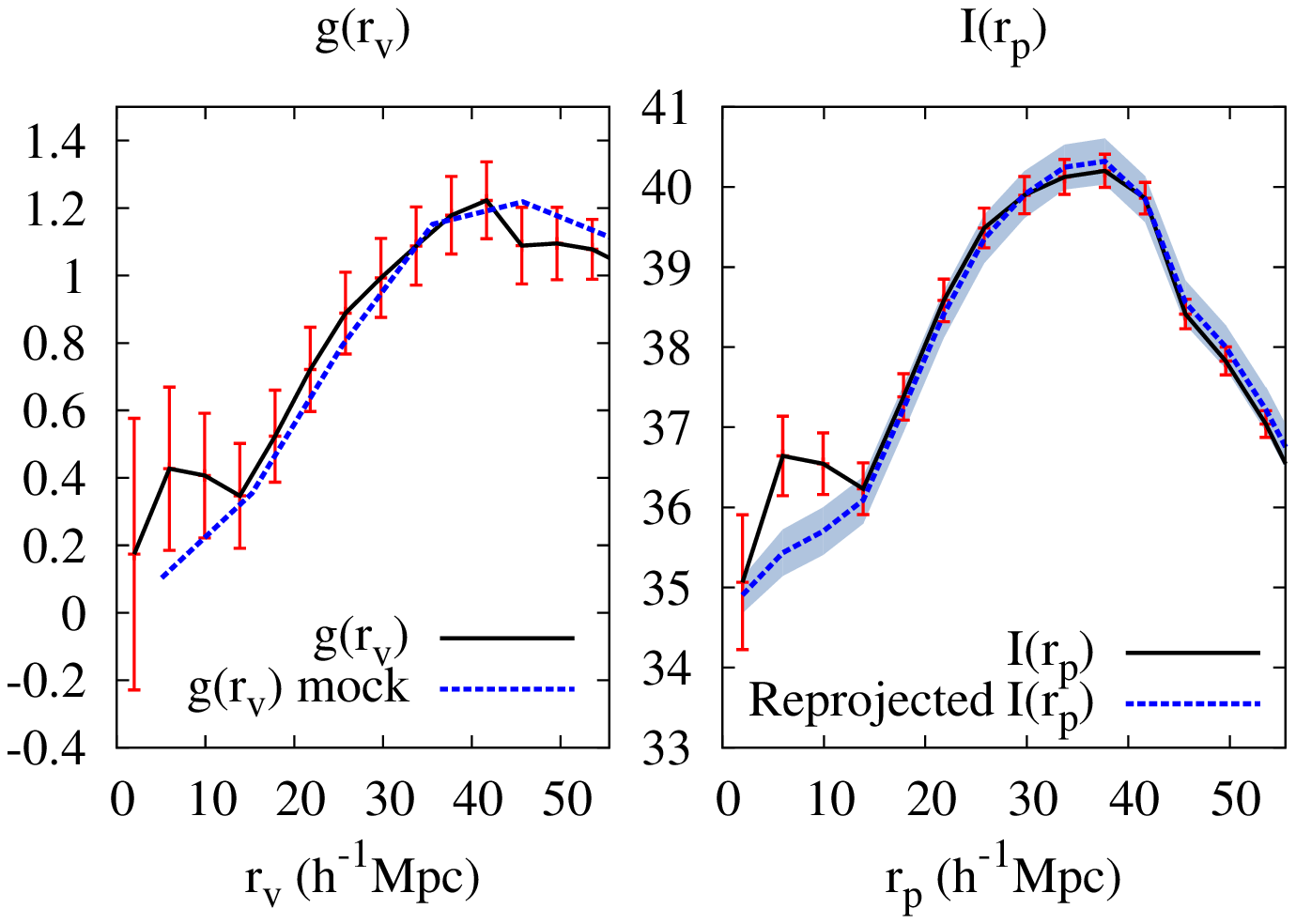,width=0.7\linewidth,clip=, angle=-90}  
\caption{Reconstruction for a 40-45 $\mathrm{h^{-1}Mpc}$ stacked void from the mock galaxy catalogue. Construction and colouring is identical to Fig. \ref{hod10-15}.\label{hod40-45}}
\end{figure}

We show in Figs \ref{hod10-15} and Fig. \ref{hod40-45} the reconstructions for stacked voids of, respectively, 10-15 $\mathrm{h^{-1}Mpc}$ and a 40-45 $\mathrm{h^{-1}Mpc}$ radii from the mock galaxy catalogue: in both cases, the reconstructed real-space stacked void profile matches the profile of the stacked void from the mock catalogue. The sanity check of the reprojection serves as an additional consistency check for the quality of the reconstruction. We notice that the first points are less precise: the error bars are higher at small radii. 

As discussed in the previous section, the reconstruction with the algorithm is more complicated at the centre, where the projection gets a major contribution from the outer shells of the sphere, resulting in an increased precision for the profile when the radius increases. As expected, this is correctly captured by the test with the reprojection, which also shows that the reconstruction is able to overcome the ill-conditioning and to recover the  real-space density profile of the stacked voids.  

The reconstruction of the spherical profile of stacked voids obtained from a mock galaxy catalogue targeted to match the properties of the SDSS DR7 sample (Figs \ref{hod10-15} and \ref{hod40-45}) has been successfully tested. The set is now ready for a first application to real data: reconstruct spherical density profiles of stacked voids from the SDSS.\\

 \begin{figure*}
\begin{tabular}{rl}
\centering
 \hspace{-4em}
  \vspace{-5pt}
\epsfig{file=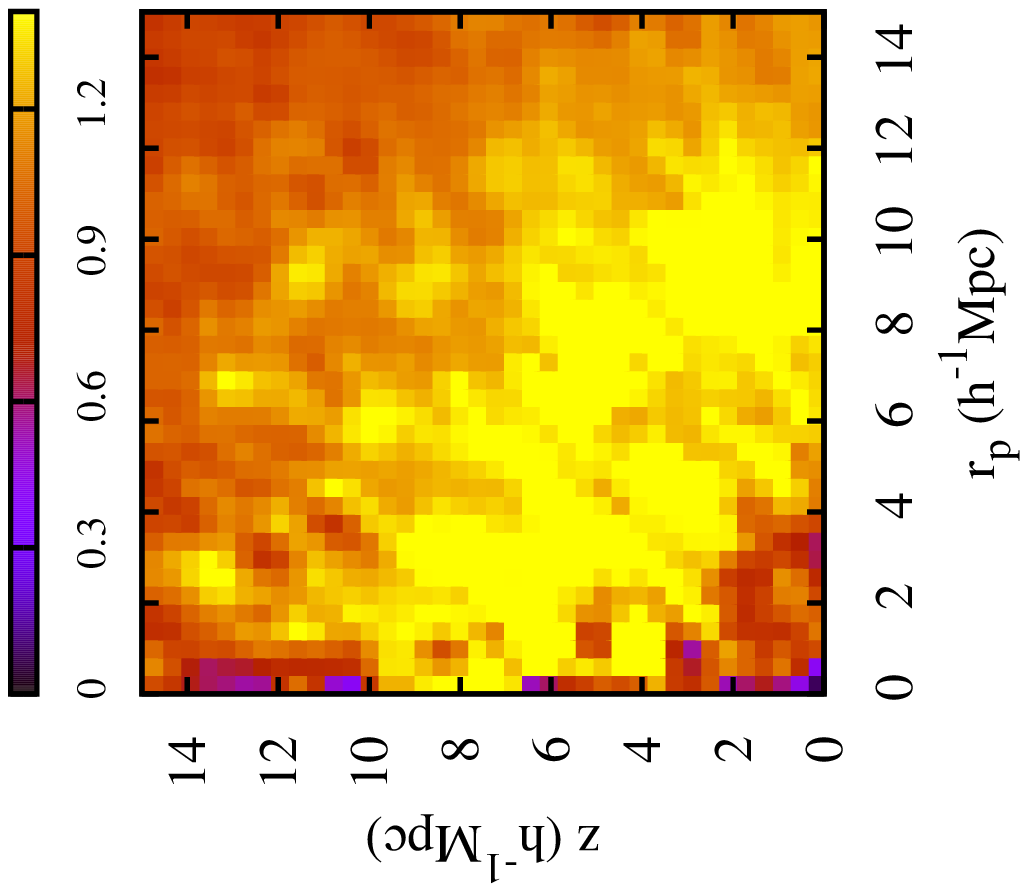,width=0.40\linewidth,clip=,angle=-90} &
\epsfig{file=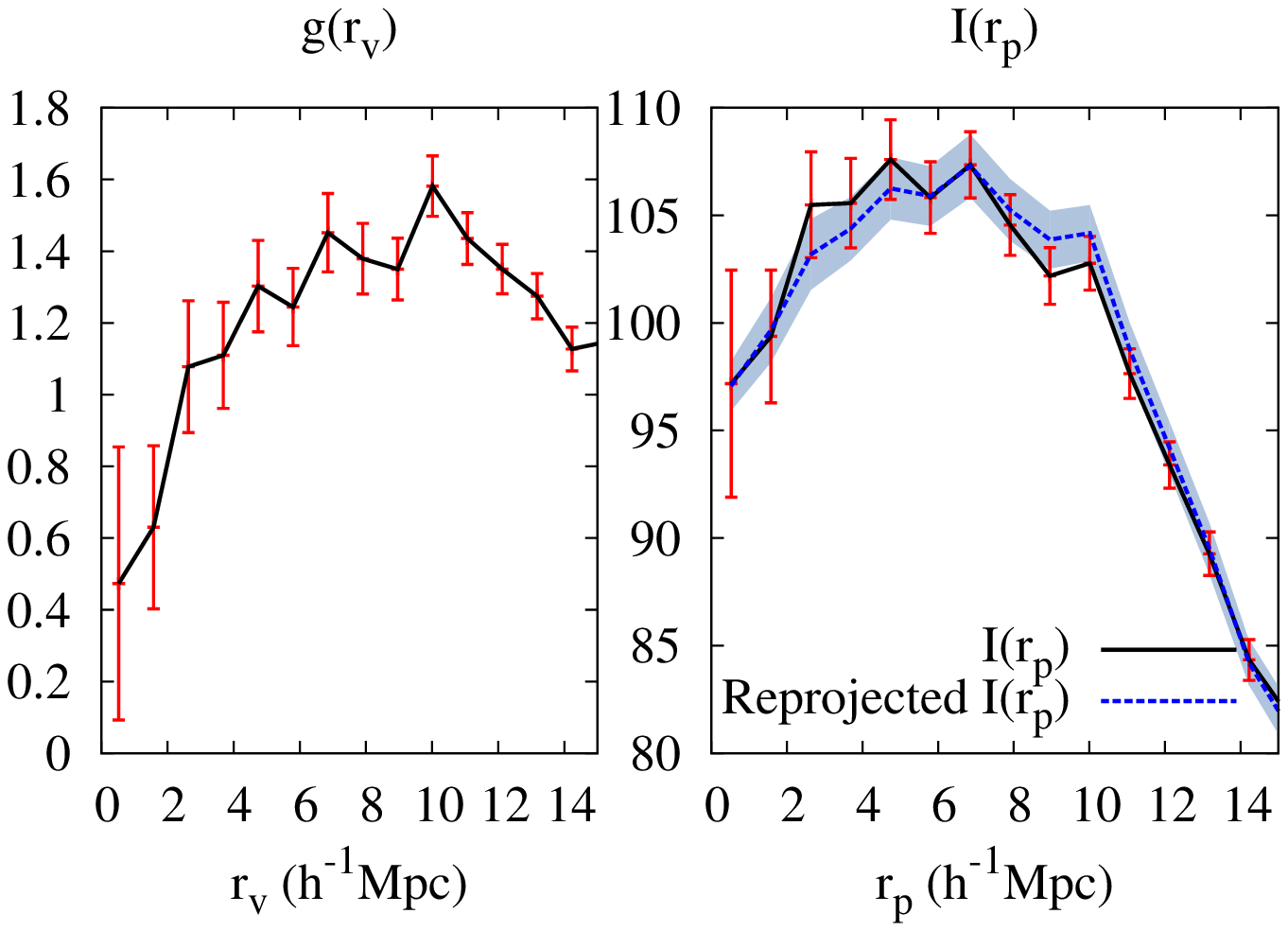,width=0.40\linewidth,clip=,angle=-90} \\
\end{tabular}
\caption{Results for a 5-15 $\mathrm{h^{-1}Mpc}$ stacked void of data set dim2: from left to right, we represent the density in redshift space $\rho(r_{\mathrm{p}},z)$, the reconstructed density $g(r_{\mathrm{v}})$ as a one-dimensional plot, and finally, the comparison between initial $I(r_{\mathrm{p}})$ (column density) and the reprojected $I(r_{\mathrm{p}})$ from the reconstruction. The light-blue bands on the right-hand plot are the errors on the reprojected $I(r_{\mathrm{p}})$ obtained by projecting the reconstructed spherical density profile $g(r_{\mathrm{v}})$. Here, we have normalized to mean density for $g$ and $\rho$ (while $I(r_{\mathrm{p}})$ units are number of galaxies per  $\mathrm{(h^{-1}Mpc)^{2}}$).\label{dim2a}}
\end{figure*}

\section{Results: density profiles for real stacked voids}

In this part we will present the results of a first application of the algorithm to the most recent real stacked voids catalogue from \cite{Sutter2012a}. The catalogue is divided in data sets based on redshift and radius of stackings. More precisely, the data sets are: dim1 (z=0.0-0.05), dim2 (z=0.05-0.1), bright1 (z=0.1-0.15), bright2 (z=0.15-0.20), lrgdim (z=0.16-0.36) and lrgbright (z=0.36-0.44). The first application shows that consistent results can be obtained from real data, for the purpose of this paper, we focus on showing the general shape of profiles in a subset of the data sets of stacked voids. 

It is clear that good reconstruction requires void stacks with a large number of voids (to converge to an isotropic stack) and galaxies (to lower Poisson noise). We will present a few first examples of real-space void profile reconstructions where these conditions hold at least approximately.

At first glance, considering the need of many voids and galaxies in the stack, we might think that stacked voids including a large range of radii for the voids sizes would give better results. This is not the case: if the range of radii for voids in the stack is too large compared to the size of the smallest voids in the stack (for example a stacking of 5-25 $\mathrm{h^{-1}Mpc}$), the wall of the stack is very thick, and the density profile noisy, since we are stacking voids with very different wall sizes and with a small common volume. Very large bins would then be undesirable since they would  mix too many void scales, the lack of rescaling in these cases would result in a very broad profile.  

Nevertheless, even if, on average the shape of voids is spherical, each void of the stack can have a different shape and a different wall thickness. Depending on the use to be done for the stacked void, it might be preferable to consider a range of radii for voids when stacking voids (instead of normalizing at the void radius). The rescaling could indeed distort the profiles and affect their use, it might thus be necessary to check whether the rescaling changes or not the properties of the stack (as discussed in Sutter et al. (2012b), where the rescaled and the non-rescaled case are compared). 

For such cases, we want to assess  the capability of the algorithm to reconstruct the real-space shape even with extreme cases -- which mean larger and possibly unscaled bins -- in the eventuality of a non-rescaling choice. As we will further discuss, the example of the 5-15 $\mathrm{h^{-1}Mpc}$ stack in Fig. \ref{dim2a} shows that the reconstruction works well even in this more extreme case: the reconstructed void has, as expected, a large wall -- the physical properties are preserved in the reconstruction. 

We finally point out that, in the eventuality of choosing to work with a range of radii for the stacks, the reconstruction algorithm remains well performing, but a balance is generally needed between too large radii stacks (to avoid poor populated voids) and too small radii stacks (to avoid mixing too many scales).

Indeed, choosing a range of radii that is too small (for example 10-12 $\mathrm{h^{-1}Mpc}$) will not be adequate in the case of real data. In such small ranges, the number of voids would be very limited, the noise on projection high and the reconstruction poor. This radius range is acceptable only for the simulation, where we have enough particles and can get a sample of 200,000 particles in a void stack with radius range of 10-12 $\mathrm{h^{-1}Mpc}$.

\begin{figure*}
\begin{tabular}{rl}
\centering
 \hspace{-4em}
  \vspace{-5pt}
\epsfig{file=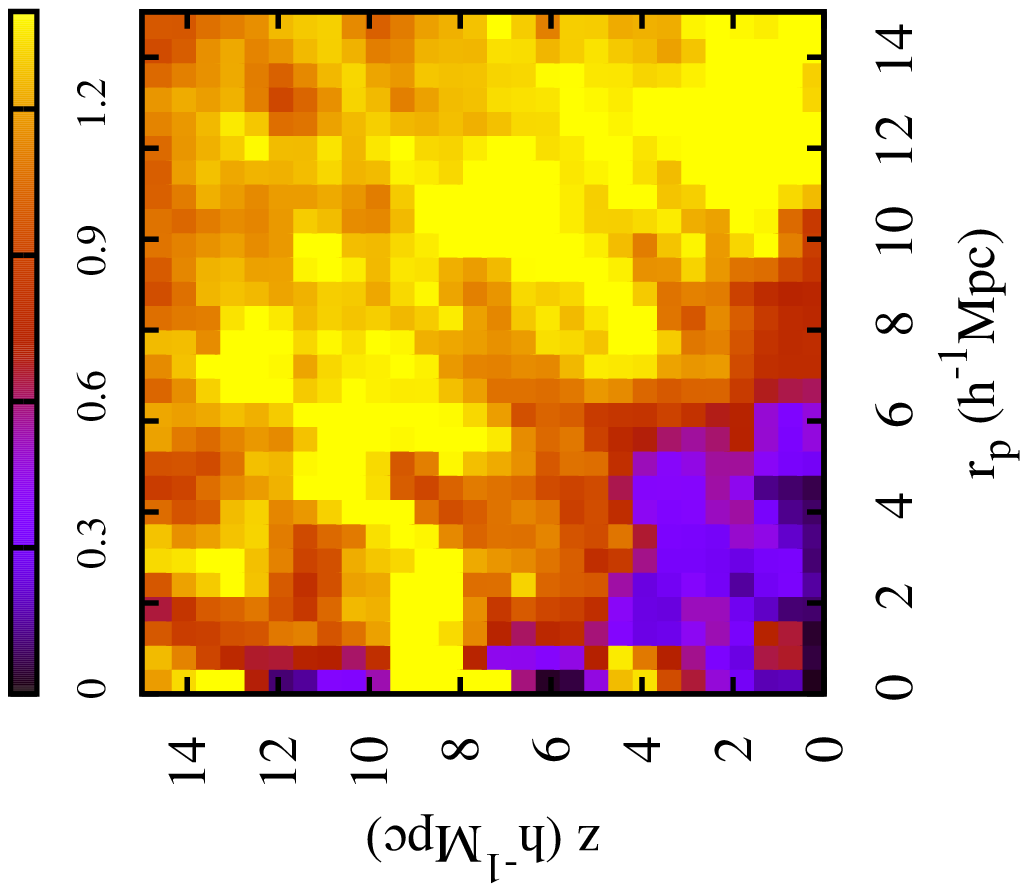,width=0.40\linewidth,clip=,angle=-90} &
\epsfig{file=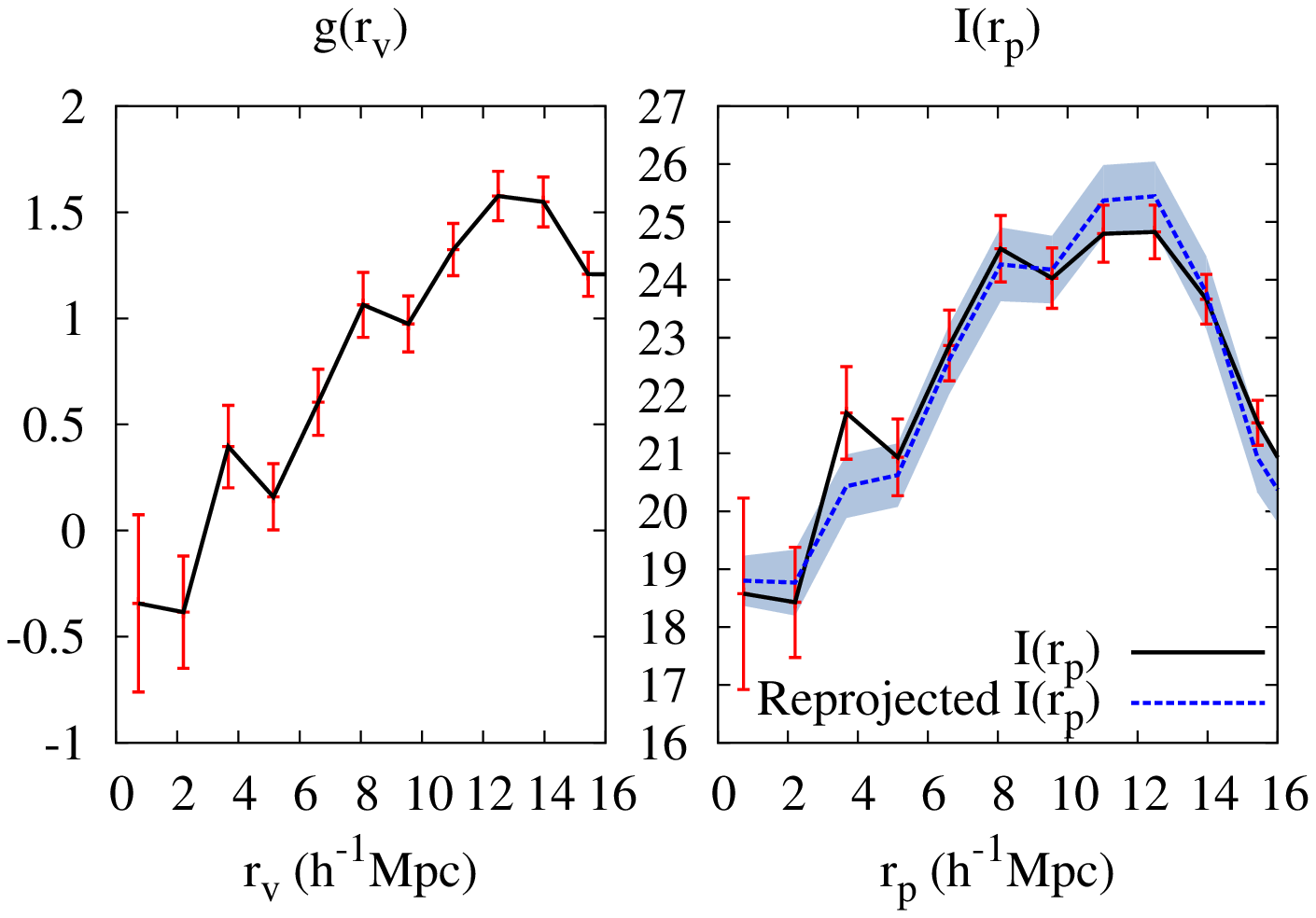,width=0.40\linewidth,clip=,angle=-90} \\
\end{tabular}
\caption{Results for a 10-15 $\mathrm{h^{-1}Mpc}$ stacked void of data set dim2: from left to right, we represent the density in redshift space $\rho(r_{\mathrm{p}},z)$, the reconstructed density $g(r_{\mathrm{v}})$ as a one-dimensional plot, and finally, the comparison between initial $I(r_{\mathrm{p}})$ (column density) and the reprojected $I(r_{\mathrm{p}})$ from the reconstruction. The light-blue bands on the right-hand plot are the errors on the reprojected $I(r_{\mathrm{p}})$ obtained by projecting the reconstructed spherical density profile $g(r_{\mathrm{v}})$. Here, we have normalized to mean density for $g$ and $\rho$ (while $I(r_{\mathrm{p}})$ units are number of galaxies per  $\mathrm{(h^{-1}Mpc)^{2}}$).\label{dim2b}}
\end{figure*}

Globally, data sets with more galaxies have lower error, so for data sets of voids with small radius (that have more voids) the error is smaller in the $I(r_{\mathrm{p}})$ and consequently also in the reconstruction $g(r_{\mathrm{v}})$. The projections of large voids have higher noise because there are less voids (and less galaxies). Furthermore, data sets at large redshift have higher noise, because less galaxies are detected at larger redshift.

So we limit the choice to low redshift and to small voids: we exclude data sets lrgbright, lrgdim and large sizes of voids (larger than 45 $\mathrm{h^{-1}Mpc}$) since they have noise-dominated projected densities. 

Finally, from the analysis of the full data set, it empirically emerges that even data sets with many voids need to have an average of at least 1000 galaxies for each void to have an acceptable signal to noise. We found that both data sets with many low populated voids and data sets with few highly populated voids have noise-dominated profiles. Only data sets well-populated in \textit{number of voids} and in \textit{number of galaxies per void} can give acceptable profiles. 

Following these considerations, to illustrate a first application of the method we have chosen stacked cosmic voids with an average of 1000 galaxies per void and (for some of them) at least 35 voids per stack. The number of voids in the stack must indeed allow the assumption of sphericity, this is why it cannot be too low. For the considered cases, the algorithm controls noise in the reconstruction and gives an acceptable spherical density profile. 

We consider the stacked voids in table \ref{tab}.
\\

\renewcommand{\tabcolsep}{0.2cm}

\begin{table}[h]
\centering 
\begin{tabular}{@{\extracolsep{\fill}} c c c c c }
\caption{Stacked cosmic voids from SDSS data. \label{tab}}
\\
\hline    
Stack radius& Redshift& Data set & Galaxies & Voids\\   
\hline                       
5-15 & 0.05-0.10 &  dim2 & 173929 &  173\\
10-15 &  0.05-0.10 & dim2 & 43527  & 41\\
20-25&  0.10-0.15& bright1 &21241& 17\\ 
25-45&  0.15-0.20& bright2 &51913 & 37\\ 
\hline    
 \end{tabular}
 \end{table}

\renewcommand{\tabcolsep}{-1cm}

In this first application, we show for each stack the distorted density profile of the stacked void in the plane $(r_{\mathrm{p}},z)$, the reconstructed spherical profile in real space (as a function of the radius of the void $r_{\mathrm{v}}$, since the profile is spherical) and the projection from which the reconstruction is done.

\begin{figure*}
\begin{tabular}{rl}
\centering
 \hspace{-4em}
  \vspace{-5pt}
\epsfig{file=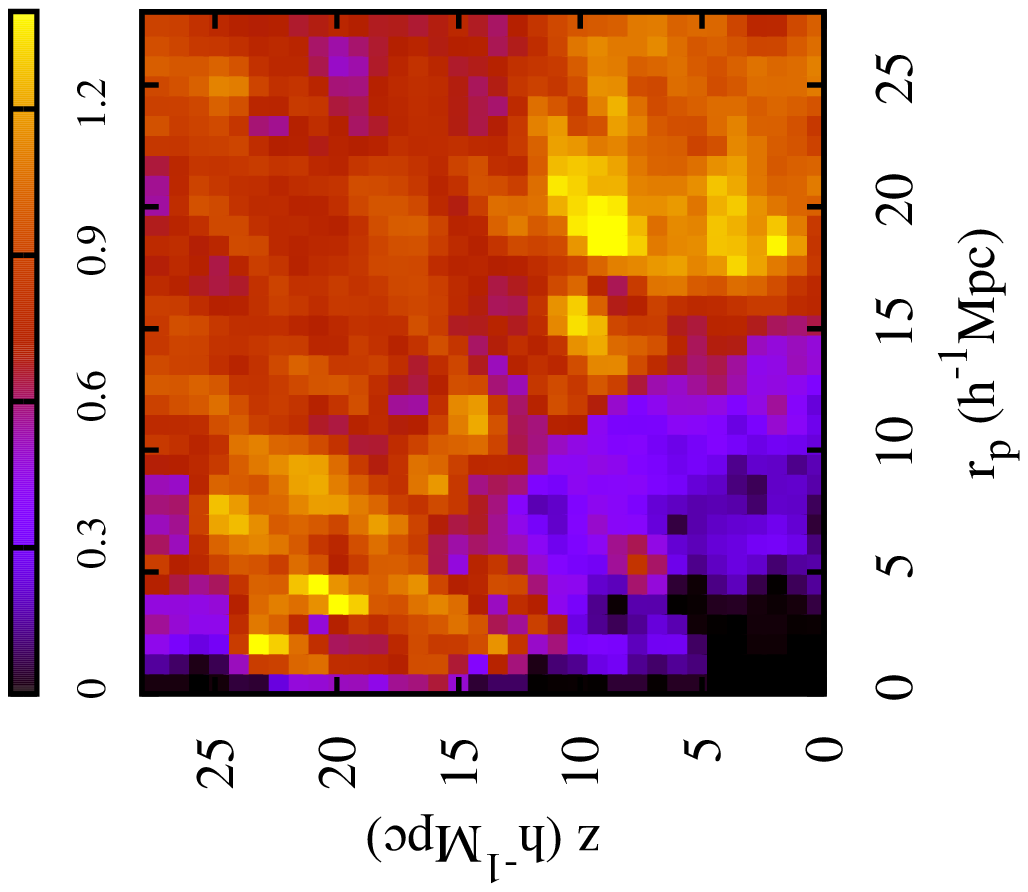,width=0.40\linewidth,clip=,angle=-90} &
\epsfig{file=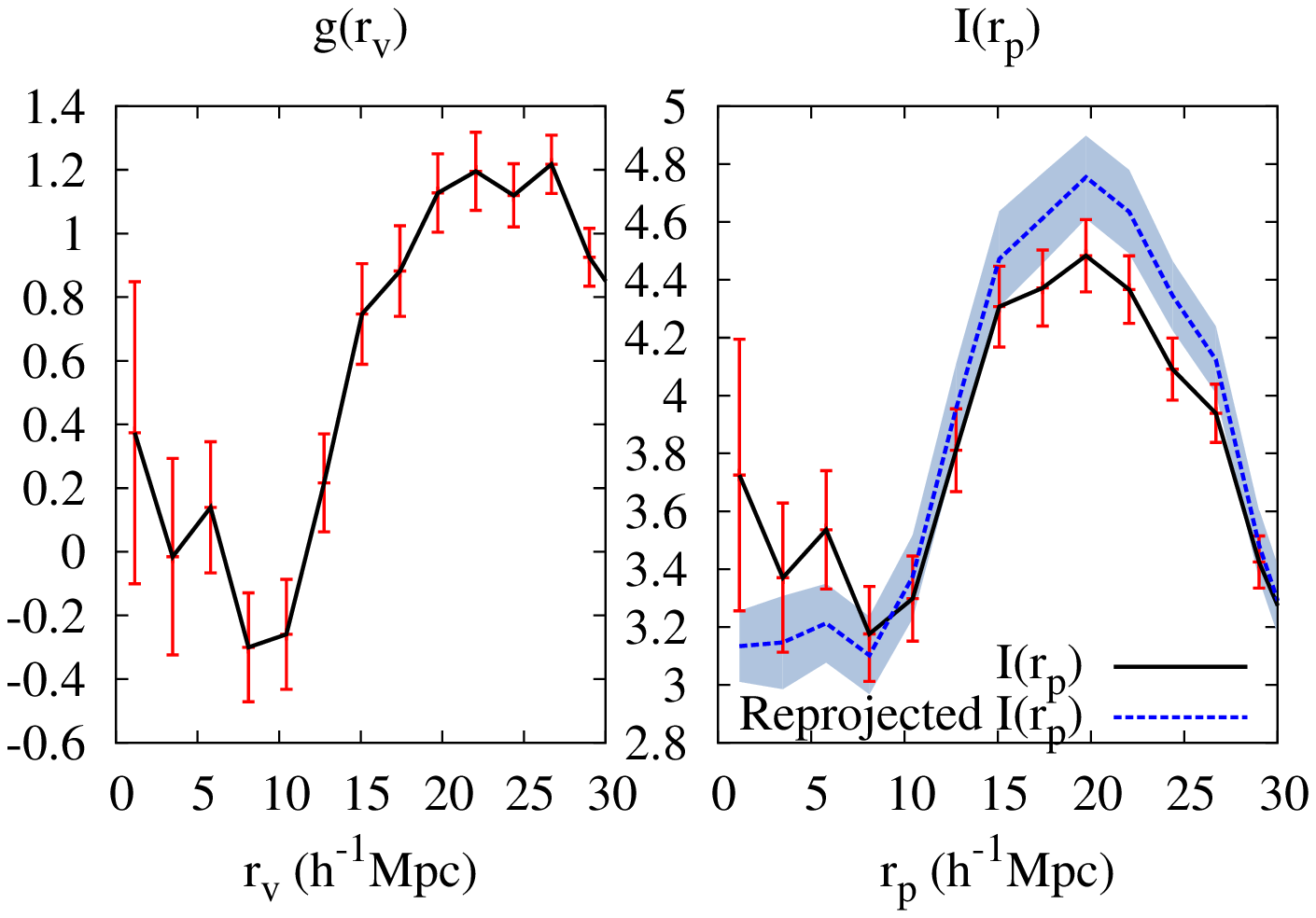,width=0.40\linewidth,clip=,angle=-90} \\
\end{tabular}
\caption{Results for a 20-25 $\mathrm{h^{-1}Mpc}$ stacked void of data set bright1: from left to right, we represent the density in redshift space $\rho(r_{\mathrm{p}},z)$, the reconstructed density $g(r_{\mathrm{v}})$ as a one-dimensional plot, and finally the comparison between initial $I(r_{\mathrm{p}})$ (column density) and the reprojected $I(r_{\mathrm{p}})$ from the reconstruction. The light-blue bands on the right-hand plot are the errors on the reprojected $I(r_{\mathrm{p}})$ obtained by projecting the reconstructed spherical density profile $g(r_{\mathrm{v}})$. Here, we have normalized to mean density for $g$ and $\rho$ (while $I(r_{\mathrm{p}})$ units are number of galaxies per  $\mathrm{(h^{-1}Mpc)^{2}}$). Low sampling leads to biases at small radii. \label{bright1}}
\end{figure*}

\begin{figure*}
\begin{tabular}{rl}
\centering
 \hspace{-4em}
  \vspace{-5pt}
\epsfig{file=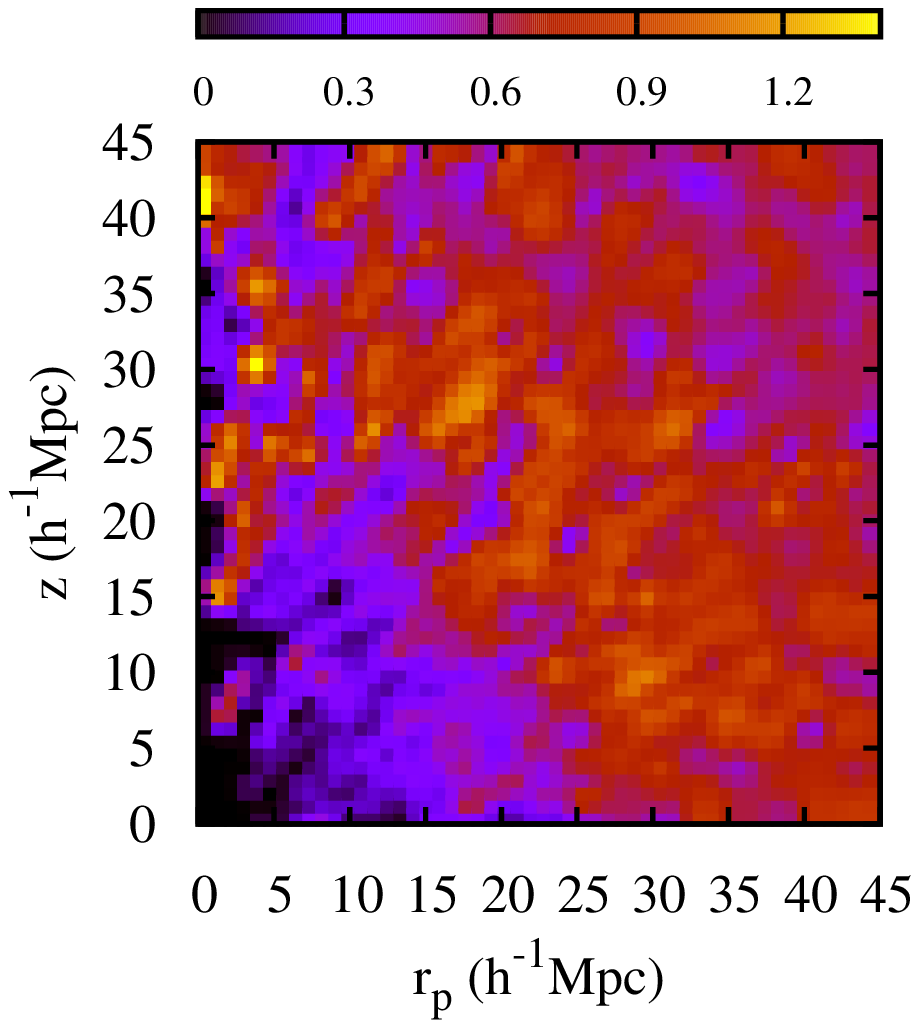,width=0.40\linewidth,clip=,angle=-90} &
\epsfig{file=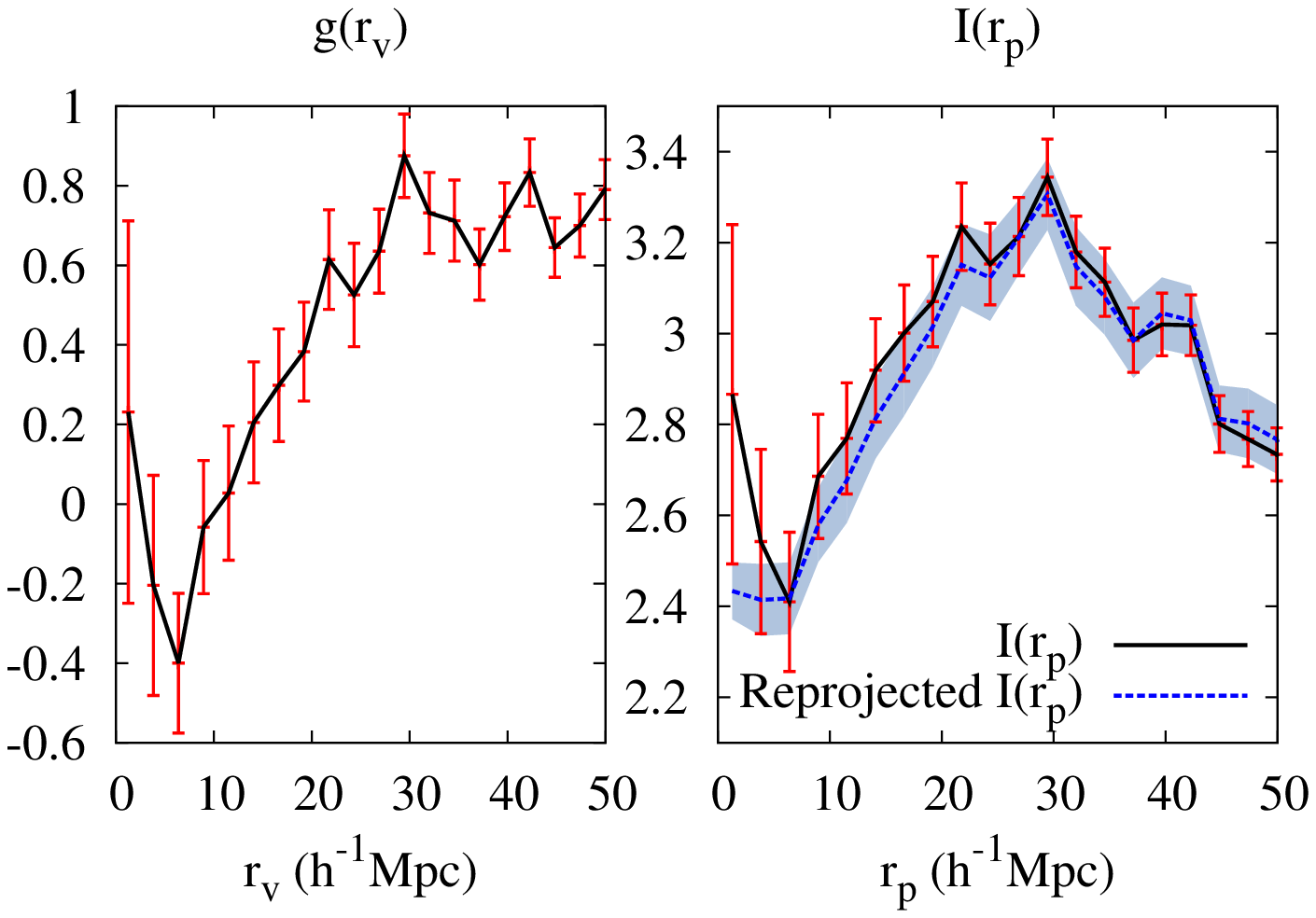,width=0.40\linewidth,clip=,angle=-90} \\
\end{tabular}
\caption{Results for a 25-45 $\mathrm{h^{-1}Mpc}$ stacked void of data set bright2: from left to right, we represent the density in redshift space $\rho(r_{\mathrm{p}},z)$, the reconstructed density $g(r_{\mathrm{v}})$ as a one-dimensional plot, and finally the comparison between initial $I(r_{\mathrm{p}})$ (column density) and the reprojected $I(r_{\mathrm{p}})$ from the reconstruction. The light-blue bands on the right-hand plot are the errors on the reprojected $I(r_{\mathrm{p}})$ obtained by projecting the reconstructed spherical density profile $g(r_{\mathrm{v}})$. Here, we have normalized to mean density for $g$ and $\rho$ (while $I(r_{\mathrm{p}})$ units are number of galaxies per  $\mathrm{(h^{-1}Mpc)^{2}}$). Low sampling leads to biases at small and large radii. \label{bright2}}
\end{figure*}

We also show, for each reconstructed profile, the reprojected density obtained from the reconstruction. In each plot of the reprojected density (right-hand plot of Figs \ref{dim2a}, \ref{dim2b}, \ref{bright1} and \ref{bright2}), the light-blue bands represent the errors on the reprojected $I(r_{\mathrm{p}})$ obtained by projecting the reconstructed spherical density profile $g(r_{\mathrm{v}})$. As discussed, we compute errors using bootstrap samples, in order to fully take into account the effects contributing to errors. The shape of the reconstructed profiles generally reaches gently the mean density. 
The reprojected density shown in Figs \ref{dim2a}, \ref{dim2b}, \ref{bright1} and \ref{bright2} generally peaks at the radius of voids since it sums all the galaxies along the line-of-sight, which at that radius includes the wall. As pointed out in Section \ref{sanity}, the comparison of the reprojected density with the initial $I(r_{\mathrm{p}})$ from data allows us to check the quality of the reconstruction, so we use the reprojected $I(r_p)$ as a diagnostic.

The reconstructions show the capability of the algorithm to obtain the spherical profile in real space even in the case of real noisy projections. All the profiles show the characteristic shape of the void: underdensity in the centre, wall and then return to mean density of the stack. As noted in the simulated stacked void, the first few points are noisier. After those initial points, the reconstruction is acceptable. 

The fact that a good reconstruction can be obtained even in the case of very noisy data is an important asset of the algorithm. The noise reduction of the Abel inversion is critical in the case of high noise in the initial projection of the stacked void, i.e. for real stacked cosmic voids. The reconstruction also validates the stacking radius, since it is now possible to check the radius of the void stacks in real space.

We now briefly comment on the profiles. For data set dim2 (Figs \ref{dim2a} and \ref{dim2b}) we choose to represent stacks with two different radii ranges for the stacking, in order to show the effect of the different, overlapping ranges on the reconstruction. The first (see Fig. \ref{dim2a}) is a stacking of voids with radii in the range 5-15 $\mathrm{h^{-1}Mpc}$, the second is a stacking of voids with radii in the range 10-15 $\mathrm{h^{-1}Mpc}$. We immediately see in the reconstruction that the wall for the stack 5-15 $\mathrm{h^{-1}Mpc}$ (see Fig. \ref{dim2a}) is thicker and the slope of the density profile is higher compared to the 10-15 $\mathrm{h^{-1}Mpc}$ stacked void (see Fig. \ref{dim2b}). This is because for the 5-15 $\mathrm{h^{-1}Mpc}$ stack, we include very small voids (with 5 $\mathrm{h^{-1}Mpc}$ of radius), so the wall starts at smaller radius. The stacking with larger bins will contain more galaxies, but the resolution for the shape of the wall will be lower and will result in a different shape. If we consider the stacking of voids with radii in the range 10-15 $\mathrm{h^{-1}Mpc}$, the compensation in the profile is narrower, since the wall does not include the wall of the voids with 5 $\mathrm{h^{-1}Mpc}$ radius.

From this, we can get two conclusions. The first is that the reconstruction of the density profile in real space correctly reflects the physical properties of the stack: we recover a thicker wall if we consider small radii voids in the stack. The second is that, if we want to extract cosmological information from stacked voids, it is necessary to be cautious in taking reasonable radius ranges for the stacks and understand well the effects of the stacking on the density profile for each application. This affects the shape of the void (and the thickness of the wall, that is the compensation). Further work with density reconstruction in real space and stacking of reconstructed profiles might help to understand the dynamics of voids and eventually study the existence of a universal profile. 

We also note that the 10-15 $\mathrm{h^{-1}Mpc}$ stacked void has slightly negative values for the first points of the profile. We did not use any prior assumption for the density to be positive, and, as observed in the case of the simulated void, the first points of the reconstruction are less precise, while the reconstruction gains in precision when the radius increase. With less galaxies, the profile loses precision in the centre: the 5-15 $\mathrm{h^{-1}Mpc}$ stack is less affected by errors because of the high number of galaxies considered (173929 galaxies, see Table \ref{tab}).
The match within the errors of the reprojected $I(r_p)$ with the density $I(r_p)$ from data (right-hand plot in Figs \ref{dim2a} and \ref{dim2b}) is a consistency check for the reconstruction of both profiles from data set dim2.

We now analyse the results for bigger voids. The stacked void from data set bright1 with radius in the range 20-25 $\mathrm{h^{-1}Mpc}$ (see Fig. \ref{bright1}), is more affected by noise, as expected because of the small number of voids. The reconstruction is noisier at small radii (lower than 10 $\mathrm{h^{-1}Mpc}$), but the algorithm still manages to reconstruct the profile. Here, the density starts increasing after 10 $\mathrm{h^{-1}Mpc}$, and its slope is higher. We observe that the inner part of the profile has density values higher than expected. This might depend on the feature of the algorithm (that gains in precision at a few points from the centre) and on the assumption of sphericity: in the case of large voids, the low sampling of galaxies might result in large asymmetries and explain the observed higher densities in the centre of voids. 

Finally, the profile of the stacked void of 25-45 $\mathrm{h^{-1}Mpc}$ of data set bright2 (Fig. \ref{bright2}) shows a lower density for the wall compared to other data sets.

We have shown as a proof of concept the first application of the algorithm to real stacked voids. The use of our algorithm with well-populated stacks of well-populated voids in the case of real data allows us to control noise in the reconstruction and to obtain the expected profile of stacked voids. In the next section, we conclude and discuss limitations and future improvements of the algorithm.

\section{Conclusions and future work}
We have presented a model-independent non-parametric algorithm to reconstruct spherical density profiles of stacked voids. We have tested the algorithm in the case of a simplistic toy model in order to illustrate the method. 

We compute the density profile in real space for a simulated stacked void. We used different methods to implement the Abel inverse with the aim of checking for consistency. The reconstruction of the density profile for the stacked void matches the profile in the simulation, showing the capability of the algorithm to obtain a reliable profile. 
Furthermore, we have tested the algorithm with a realistic mock galaxy catalogue mimicking data from the SDSS DR7. The mocks provide a validation of the algorithm in the case of scenarios with realistic signal to noise, further enhancing its reliability for the application to real data.

Finally, we showed a first application of the algorithm to real data and obtained the spherical density profile of real, well-populated stacked voids from the catalogue of \cite{Sutter2012a}. We set some constraints on the number of galaxies needed for each void of the stack (at least 1000 galaxies per void) and on the number of voids of the stack necessary to allow the algorithm to overcome noise (35 voids). We have shown the capability of the algorithm to control noise in the reconstruction of the void density profile in real space solely assuming (asymptotic) sphericity, i.e. without introducing a prior on cosmological parameters or a dynamical model of voids. 

The main limitation of the algorithm remains the high noise in the projection for data sets at high redshift and for large voids. Introducing reasonable priors may improve the reconstruction at the expense of giving up some of the explicit model independence. 
In the reconstructed stacked void density profiles, the shape and value of the overdensity of the wall (the compensation) has an important role in understanding the physics of the void and is another factor to be investigated in future work.  The reconstructed density $g(r_{\mathrm{v}})$ might allow in future to discriminate between different cosmological models. 

This first application of the algorithm on real voids is a proof of concept, the first step to a better understanding of the shape of voids. It is important to determine the reason of these differences in the shape of voids, that might depend on many factors (on the radius, physics and evolution of the stacked void). \cite{Lavaux2012} and \cite{Sutter2012b} suggested the presence of a common profile for stacked voids of different radii. The reconstruction of density profiles in real space offers the possibility to analyse this claim in observations and we assess for further work its detailed investigation.  A future possible improvement of the algorithm would be the rescaling of the reconstructed profile for different sizes of voids to obtain statistical properties of profiles.

As for future applications, since the Alcock-Paczy\'{n}ski test relies on the difference between the shape of void in redshift space and in real space to measure the expansion of the Universe, the cosmological-independent shape of the voids density profile in real space can help to reduce the systematic error in the test \citep{Sutter2012b}: it would give the exact shape of the void to compare with the distorted shape of the void in redshift-space data. Furthermore, a complete knowledge of the real density profile of voids will allow studying their evolution without being affected by redshift distortions. Among other applications, we will consider the reconstruction of the expansion of voids and their velocity profile. 

Finally, \cite{Verde2013} argued that a local cosmological-independent measure of the Hubble parameter (that can be provided by the Alcock-Paczy\'{n}ski test) may help understanding the discrepancy suggested by recent data for the value of $H_{0}$ (see \cite{Riess1998}, \cite{Perlmutter1999}, \cite{Planck2013} but also discussions in \cite{Fleury2013} and \cite{Marra2013}). Models of modified gravity (such as fifth force models) and dark energy (\textit{e. g.} \cite{Clampitt2013}, \cite{Spolyar2013}, \cite{Sutter2014}) could be constrained with our algorithm: considering the shape of the density profiles on simulations with the models and the shape of profiles obtained applying our algorithm to observational data, we could discriminate between such models.  The reconstruction method does not make any cosmological assumption about the model, thus the density profile reconstruction of stacked voids in real space opens the way to better constrain the value of the Hubble constant and eventually cosmological models and new physics on current and future data sets such as the \textit{Euclid} survey \citep{Laureijs2011}.

\section*{Acknowledgements}

AP thanks Nico Hamaus for useful discussions. The authors thank the anonymous referee for the pertinence of her/his comments that helped to clarify some points and to generally improve the paper. The authors acknowledge support from NSF Grant NSF AST 09- 08693 ARRA. 
PMS and BDW acknowledge support from NSF Grant AST-0908902. BDW is partially supported by a senior Excellence Chair by the Agence Nationale de Recherche (ANR-10-CEXC-004-01).
BDW and AP acknowledge support from BDW's Chaire Internationale in Theoretical Cosmology at the Universit\'{e} Pierre et Marie Curie. GL acknowledges support from CITA National Fellowship and financial support from the Government of Canada Post-Doctoral Research Fellowship. Research at Perimeter Institute is supported by the Government of Canada through Industry Canada and by the Province of Ontario through the Ministry of Research and Innovation.

\bibliographystyle{mn2e}
\bibliography{biblio_new_noarxiv}

\begin{thebibliography}{}

\bibitem[\protect\citeauthoryear{Abel}{Abel}{1988}]{Abel}
Abel N.~H.,  1842, reprint 1988, Oeuvres Completes.
SEd. L. Sylow and S. Lie, New York: Johnson Reprint Corp.

\bibitem[\protect\citeauthoryear{}{Akaike}{1974}]{Akaike1974}
Akaike H.,  1974, IEEE Transactions on Automatic Control, 19, 716

\bibitem[\protect\citeauthoryear{}{{Alcock} \& {Paczynski}}{1979}]{Alcock1979}
{Alcock} C.,  {Paczynski} B.,  1979, \nat, 281, 358

\bibitem[\protect\citeauthoryear{}{{Andrae} et~al.}{2010}]{Andrae2010}
{Andrae} R.,  {Schulze-Hartung} T.,    {Melchior} P.,  2010, ArXiv e-prints,
  {arXiv:1012.3754}

\bibitem[\protect\citeauthoryear{}{{Aragon-Calvo} \&
  {Szalay}}{2013}]{Aragon-Calvo2013}
{Aragon-Calvo} M.~A.,  {Szalay} A.~S.,  2013, \mnras, 428, 3409

\bibitem[\protect\citeauthoryear{}{Blas et~al.}{2011}]{Blas2011}
Blas D.,  Lesgourgues J.,    Tram T.,  2011, Journal of Cosmology and
  Astroparticle Physics, 2011, 034

\bibitem[\protect\citeauthoryear{}{{Bond} et~al.}{1996}]{Bond1996}
{Bond} J.~R.,  {Kofman} L.,    {Pogosyan} D.,  1996, \nat, 380, 603

\bibitem[\protect\citeauthoryear{}{{Bos} et~al.}{2012}]{Bos2012}
{Bos} E.~G.~P.,  {van de Weygaert} R.,  {Ruwen} J.,  {Dolag} K.,    {Pettorino}
  V.,  2012, ArXiv e-prints, {arXiv:1211.3249}

\bibitem[\protect\citeauthoryear{Bracewell}{Bracewell}{1999}]{Bracewell}
Bracewell R.,  1999, The Fourier Transform and Its Applications, 3rd ed..
McGraw-Hill

\bibitem[\protect\citeauthoryear{Burnham \& Anderson}{Burnham \&
  Anderson}{2002}]{Akaike2002}
Burnham Anderson 2002, Model Selection and Multimodel Inference: A Practical
  Information-Theoretic Approach (2nd ed.).
Springer-Verlag

\bibitem[\protect\citeauthoryear{}{{Ceccarelli} et~al.}{2006}]{Ceccarelli2006}
{Ceccarelli} L.,  {Padilla} N.~D.,  {Valotto} C.,    {Lambas} D.~G.,  2006,
  \mnras, 373, 1440

\bibitem[\protect\citeauthoryear{}{{Chincarini}}{2013}]{Chincarini2013}
{Chincarini} G.,  2013, ArXiv e-prints, {arXiv:1305.2893}

\bibitem[\protect\citeauthoryear{}{{Clampitt} et~al.}{2013}]{Clampitt2013}
{Clampitt} J.,  {Cai} Y.-C.,    {Li} B.,  2013, \mnras, 431, 749

\bibitem[\protect\citeauthoryear{}{Crocce et~al.}{2006}]{Crocce2006}
Crocce M.,  Pueblas S.,    Scoccimarro R.,  2006, Monthly Notices of the Royal
  Astronomical Society, 373, 369

\bibitem[\protect\citeauthoryear{}{{de Lapparent}
  et~al.}{1986}]{deLapparent1986}
{de Lapparent} V.,  {Geller} M.~J.,    {Huchra} J.~P.,  1986, \apjl, 302, L1

\bibitem[\protect\citeauthoryear{}{Durret et~al.}{1999}]{Durret1999}
Durret F.,  Gerbal D.,  Lobo C.,    Pichon C.,  1999, Astronomy and
  Astrophysics, 343, 760

\bibitem[\protect\citeauthoryear{}{{Fleury} et~al.}{2013}]{Fleury2013}
{Fleury} P.,  {Dupuy} H.,    {Uzan} J.-P.,  2013, Physical Review Letters, 111,
  091302

\bibitem[\protect\citeauthoryear{}{{Gregory} \& {Thompson}}{1978}]{Gregory1978}
{Gregory} S.~A.,  {Thompson} L.~A.,  1978, \apj, 222, 784

\bibitem[\protect\citeauthoryear{}{Hamaus et~al.}{2014}]{Hamaus2014}
Hamaus N.,  Sutter P.,    Wandelt B.~D.,  2014, arXiv preprint arXiv:1403.5499

\bibitem[\protect\citeauthoryear{}{{Hamilton}}{1998}]{Hamilton1998}
{Hamilton} A.~J.~S.,  1998, in {Hamilton} D.,  ed., The Evolving Universe
  Vol.~231 of Astrophysics and Space Science Library, {Linear Redshift
  Distortions: a Review}.
p.~185

\bibitem[\protect\citeauthoryear{}{{J{\~o}eveer} et~al.}{1978}]{Joeveer1978}
{J{\~o}eveer} M.,  {Einasto} J.,    {Tago} E.,  1978, \mnras, 185, 357

\bibitem[\protect\citeauthoryear{}{{Kirshner} et~al.}{1981}]{Kirshner1981}
{Kirshner} R.~P.,  {Oemler} Jr. A.,  {Schechter} P.~L.,    {Shectman} S.~A.,
  1981, \apjl, 248, L57

\bibitem[\protect\citeauthoryear{}{{Laureijs} et~al.}{2011}]{Laureijs2011}
{Laureijs} R.,  {Amiaux} J.,  {Arduini} S.,  {Augu{\`e}res} J.~.,  {Brinchmann}
  J.,  {Cole} R.,  {Cropper} M.,  {Dabin} C.,  {Duvet} L.,  {Ealet} A.,    et
  al. 2011, ArXiv e-prints, {arXiv:1110.3193}

\bibitem[\protect\citeauthoryear{}{{Lavaux} \& {Wandelt}}{2012}]{Lavaux2012}
{Lavaux} G.,  {Wandelt} B.~D.,  2012, \apj, 754, 109

\bibitem[\protect\citeauthoryear{}{{Lee} \& {Park}}{2009}]{Lee2009}
{Lee} J.,  {Park} D.,  2009, \apjl, 696, L10

\bibitem[\protect\citeauthoryear{}{{Li} et~al.}{2007}]{Li2007}
{Li} X.-F.,  {Huang} L.,    {Huang} Y.,  2007, Journal of Physics A
  Mathematical General, 40, 347

\bibitem[\protect\citeauthoryear{}{{Marra} et~al.}{2013}]{Marra2013}
{Marra} V.,  {Amendola} L.,  {Sawicki} I.,    {Valkenburg} W.,  2013, Physical
  Review Letters, 110, 241305

\bibitem[\protect\citeauthoryear{}{{Neyrinck}}{2008}]{Neyrinck2008}
{Neyrinck} M.~C.,  2008, \mnras, 386, 2101

\bibitem[\protect\citeauthoryear{}{Patiri et~al.}{2012}]{Patiri2012}
Patiri S.~G.,  Betancort-Rijo J.,    Prada F.,  2012, Astronomy \&
  Astrophysics, 541, L4

\bibitem[\protect\citeauthoryear{}{{Perlmutter} et~al.}{1999}]{Perlmutter1999}
{Perlmutter} S.,  {Aldering} G.,    {Goldhaber} G. e.~a.,  1999, \apj, 517, 565

\bibitem[\protect\citeauthoryear{}{{Planck Collaboration}}{2013}]{Planck2013}
{Planck Collaboration} 2013, ArXiv e-prints, {arXiv:1303.5076}

\bibitem[\protect\citeauthoryear{}{{Riess} et~al.}{1998}]{Riess1998}
{Riess} A.~G.,  {Filippenko} A.~V.,    {Challis} P. e.~a.,  1998, \aj, 116,
  1009

\bibitem[\protect\citeauthoryear{}{{Ryden}}{1995}]{Ryden1995}
{Ryden} B.~S.,  1995, \apj, 452, 25

\bibitem[\protect\citeauthoryear{}{{Ryden} \& {Melott}}{1996}]{Ryden1996}
{Ryden} B.~S.,  {Melott} A.~L.,  1996, \apj, 470, 160

\bibitem[\protect\citeauthoryear{}{{Spolyar} et~al.}{2013}]{Spolyar2013}
{Spolyar} D.,  {Sahl{\'e}n} M.,    {Silk} J.,  2013, ArXiv e-prints,
  {arXiv:1304.5239}

\bibitem[\protect\citeauthoryear{}{Strauss et~al.}{2002}]{Strauss2002}
Strauss M.~A.,  Weinberg D.~H.,  Lupton R.~H.,  Narayanan V.~K.,  Annis J.,
  Bernardi M.,  Blanton M.,  Burles S.,  Connolly A.,  Dalcanton J.,    et~al.,
  2002, The Astronomical Journal, 124, 1810

\bibitem[\protect\citeauthoryear{}{Sutter et~al.}{2014a}]{Sutter2014a}
Sutter P.,  Lavaux G.,  Hamaus N.,  Pisani A.,  Wandelt B.~D.,  Warren M.~S.,
  Villaescusa-Navarro F.,  Zivick P.,  Mao Q.,    Thompson B.~B.,  2014a, arXiv
  preprint arXiv:1406.1191

\bibitem[\protect\citeauthoryear{}{Sutter et~al.}{2013}]{Sutter2013}
Sutter P.,  Lavaux G.,  Wandelt B.~D.,  Hamaus N.,  Weinberg D.~H.,    Warren
  M.~S.,  2013, arXiv preprint arXiv:1309.5087

\bibitem[\protect\citeauthoryear{}{Sutter et~al.}{2014b}]{Sutter2014}
Sutter P.~M.,  Carlesi E.,  Wandelt B.~D.,    Knebe A.,  2014b, arXiv preprint
  arXiv:1406.0511

\bibitem[\protect\citeauthoryear{}{{Sutter} et~al.}{2012a}]{Sutter2012b}
{Sutter} P.~M.,  {Lavaux} G.,  {Wandelt} B.~D.,    {Weinberg} D.~H.,  2012a,
  \apj, 761, 187

\bibitem[\protect\citeauthoryear{}{{Sutter} et~al.}{2012b}]{Sutter2012a}
{Sutter} P.~M.,  {Lavaux} G.,  {Wandelt} B.~D.,    {Weinberg} D.~H.,  2012b,
  \apj, 761, 44

\bibitem[\protect\citeauthoryear{}{{Thompson} \&
  {Gregory}}{2011}]{Thompson2011}
{Thompson} L.~A.,  {Gregory} S.~A.,  2011, ArXiv e-prints, {arXiv:1109.1268}

\bibitem[\protect\citeauthoryear{}{{Tinker} et~al.}{2006}]{Tinker2006}
{Tinker} J.~L.,  {Weinberg} D.~H.,    {Zheng} Z.,  2006, \mnras, 368, 85,
  {arXiv:astro-ph/0501029}

\bibitem[\protect\citeauthoryear{}{{Tully} \& {Fisher}}{1978}]{Tully1978}
{Tully} R.~B.,  {Fisher} J.~R.,  1978, in {Longair} M.~S.,  {Einasto} J.,  eds,
  Large Scale Structures in the Universe Vol.~79 of IAU Symposium, {Nearby
  small groups of galaxies}.
pp 31--45

\bibitem[\protect\citeauthoryear{}{{Verde} et~al.}{2013}]{Verde2013}
{Verde} L.,  {Jimenez} R.,    {Feeney} S.,  2013, Physics of the Dark Universe,
  2, 65

\bibitem[\protect\citeauthoryear{}{Zehavi et~al.}{2011}]{Zehavi2011}
Zehavi I.,  Zheng Z.,  Weinberg D.~H.,  Blanton M.~R.,  Bahcall N.~A.,  Berlind
  A.~A.,  Brinkmann J.,  Frieman J.~A.,  Gunn J.~E.,  Lupton R.~H.,    et~al.,
  2011, The Astrophysical Journal, 736, 59

\bibitem[\protect\citeauthoryear{}{Zheng et~al.}{2007}]{Zheng2007}
Zheng Z.,  Coil A.~L.,    Zehavi I.,  2007, The Astrophysical Journal, 667, 760

\end{thebibliography}

\bsp

\label{lastpage}

\end{document}